\documentclass[intlimits,twoside,a4paper]{article}

\usepackage[eqsecnum]{cmpj3}

\usepackage[cp1251]{inputenc}

\usepackage{bm}

\newcommand{\ps}{\,\textrm{ps}}
\newcommand{\fs}{\,\textrm{fs}}

\issue{2022}{25}{2}{23201}
\doinumber{10.5488/CMP.25.23201}

\title[Curcumin in water]
{Structural aspects of the clustering of curcumin molecules in water. Molecular dynamics
computer simulation study}
\author[T. Patsahan, O. Pizio]{T. Patsahan\orcid{0000-0002-7870-2219}\refaddr{label1},
	O. Pizio\orcid{0000-0001-8333-4652}\refaddr{label2}
\thanks{Corresponding author: \email{oapizio@gmail.com}.}}
\addresses{
	\addr{label1} Institute for Condensed Matter Physics of the National Academy of Sciences of Ukraine,
	1 Svientsitskii St., 79011 Lviv, Ukraine
	\addr{label2} Instituto de Qu\'{i}mica, Universidad Nacional Aut\'{o}noma de M\'{e}xico,
	Circuito Exterior, 04510, Cd. Mx., M\'{e}xico
}

\Keywords{curcumin, united atom model, molecular dynamics, water, clusters}
\date{Received April 3, 2022, in final form May 10, 2022}
\begin{document}

\maketitle

\begin{abstract}
 We explore clustering of curcumin molecules in water by using
 the OPLS-UA model for the enol conformer of curcumin
 (J. Mol. Liq., \textbf{223}, 707, 2016) and the SPC-E water model.
 With this purpose, solutions of  2, 4, 8, 12, 16 and 20 curcumin molecules
 in 3000 water molecules are studied by using extensive molecular dynamics
 computer simulations. Radial distributions for the centers of
 mass of curcumin molecules are evaluated and the running coordination
 numbers are analyzed. The formation of clusters on time is elucidated.
 The internal structure of molecules within the cluster is described by using
 radial distributions of the elements of the curcumin molecule, the orientation
 descriptors, the order parameter and the radius of gyration. The self-diffusion coefficient
 of solute molecules in clusters is evaluated. The distribution of water species 
 around clusters is described in detail. A comparison of our findings with computer simulation results
 of other authors is performed. A possibility to relate predictions of the model 
 with experimental observations is discussed.
\printkeywords
%
%
\end{abstract}

\section{Introduction}

Curcumin is well known as a spice and natural coloring agent. Its pharmacological activity 
resulting in therapeutic applications were the subject of very many experimental studies 
during the recent decades, see e.g.,~\cite{ghosh1,kumar1,luthra1,mehanny1,ngo1}.
Experimental investigations were accompanied by the applications of methods 
of quantum chemistry in this area of research. 
However, from a theoretical modelling perspective there is a huge number of variables 
involved in the experiments. Consequently, there is much room for the development of 
adequate modelling and verification of theoretical predictions against 
experimental data~\cite{wright1}.

The studies of solutions with curcumin solutes by using computer simulation techniques
were initiated quite recently
\cite{ngo2,Suhartanto-2012,Varghese-2009,Wallace-2013,Samanta-2013,Hazra-2014,Yadav-2014,Parameswari-2015,Priyadarsini-2009}.
In particular, modelling of the curcumin molecule force field  from quantum chemical (QC)
calculations, mainly within the B3LYP method and  using different versions of Gaussian software,
was undertaken in~\cite{Samanta-2013,Hazra-2014,bonab,pereira}.
Justification for the development of the simpler, OPLS-united atom (UA) model
(using the OPLS library~\cite{OPLS-1996})  for curcumin molecule 
from our laboratory was provided in~\cite{Ilny-2016}. 
The model was tested in vacuum and in water using classical molecular
dynamics simulations~\cite{Ilny-2016}.

The present work is a continuation of our research of 
a single curcumin molecule in water, methanol and dimethylsulfoxide  in the
framework of molecular dynamics computer simulation methodology~\cite{Ilny-2016,Pat-2017}.
In contrast to the previous studies of many authors, we make one important step
forward. Namely, we focus on the systems with a larger number of
curcumin molecules ($N_\text{cur}=2$, $4$, $8$, $12$, $16$ and $20$)
rather than a monomer.
Thus, the effects of interaction between solute molecules compete with
solute--solvent interaction. This kind of an appealing setup
was realized only in \cite{Hazra-2014,bonab,pereira}, up to our best knowledge.
It is important to mention that the optimization of the curcumin dimer structure 
within the QC model was undertaken in \cite{pereira}.
Here, however, we follow the development used in \cite{Hazra-2014}.
In close similarity to that work, we focus on the systems with a few
curcumin solute molecules in water. It is known that the curcumin molecule
is hydrophobic due to the presence of phenyl rings. Hence, the solute
is characterized by a very low solubility in water. Water solvent promotes 
hydrophobic association of curcumin molecules which leads to the formation of 
clusters. The study of this kind of phenomenon is the principal focus of the present work.  

\section{Model and simulation details}

Molecular dynamics computer simulations of curcumin molecules in water was performed
in the isothermal-isobaric (NPT) ensemble at a temperature of $298.15$~K and at $1$~bar.
The GROMACS simulation software \cite{GROMACS} version 5.1.4 was used. 

\begin{figure}[!t]
\begin{center}
\includegraphics[width=0.5\textwidth,angle=0,clip=true]{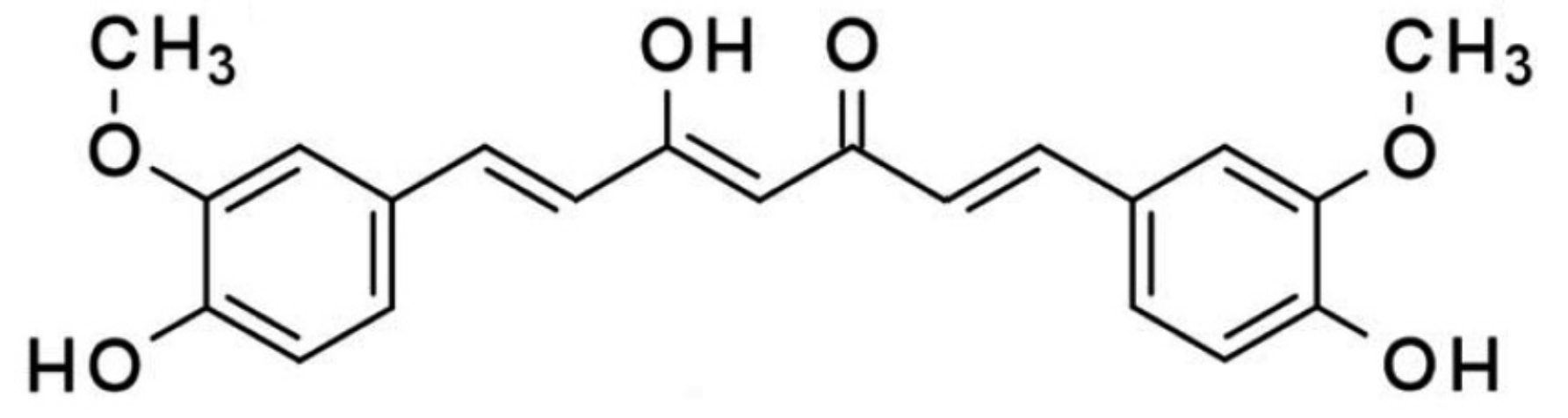}
\caption{\label{fig_enol} Chemical structure of the enol form of curcumin molecule.}
\end{center}
\end{figure}

Water is described in the framework of popular and well tested SPC-E model~\cite{SPCE_model}. 
On the other hand, the model developed by using the OPLS-UA force field is used for 
curcumin molecule ~\cite{Ilny-2016}. Recently, the model was tested by considering a single
curcumin molecule in water, methanol and dimethylsulfoxide~\cite{Pat-2017}.
Similarly to our previous investigations, in the present work
 we restrict our attention to the enol tautomer of curcumin.
This structure is dominant in solids and in various solvents, see e.g., \cite{slabber,kolev}.
The chemical structure of the molecule is shown in figure~\ref{fig_enol}.

\begin{figure}[!t]
\begin{center}
\includegraphics[width=0.5\textwidth,angle=0,clip=true]{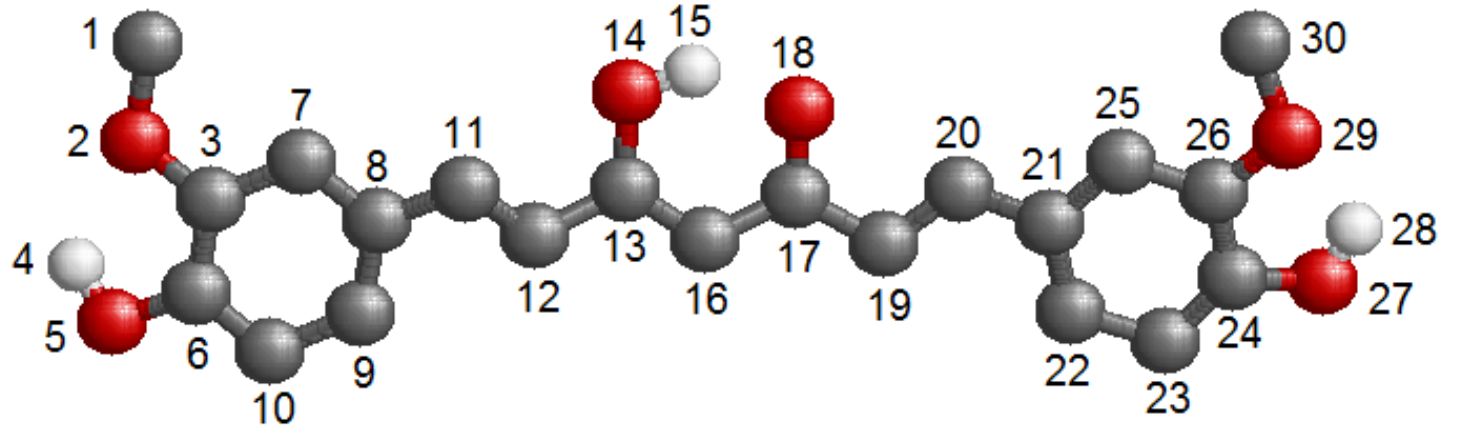}
\caption{\label{fig_curc}(Colour online) Schematic representation for the united-atom
curcumin model with sites numbering. Carbon groups are shown as dark gray spheres,
oxygens --- as red spheres, hydrogens --- as small light-gray spheres.}
\end{center}
\end{figure}

The ``ball-and-stick'' schematic representation of the molecule 
was already presented  in \cite{Ilny-2016,Pat-2017}. Still, we show
it here in figure~\ref{fig_curc}, for the sake of convenience of the reader.

All the parameters of the force field can be found in the supporting information 
file to Ref.~\cite{Ilny-2016}. 
Slight modification of the original model was undertaken, however. 
In the study of the original model, we observed quite large fluctuations of the bond length 
between oxygen and hydrogen atoms in the hydroxyl groups H4-O5, H15-O14 as well as in  H28-O27. 
In order to avoid this nonphysical behavior, in the present study
these bonds are considered as rigid with the length 0.95~\AA  ~and accounted for via constraints. 
The LINCS algorithm is used.

Technical details of the simulation procedure are as follows. 
The geometric combination rules were used to determine the parameters for the cross interactions 
(rule~3 of GROMACS software). To evaluate the Coulomb interaction contributions, the particle
mesh Ewald procedure was used (fourth-order spline interpolation and grid spacing for the fast Fourier
transform equal to $0.12$~nm). 
The cut-off distance both for Coulomb and Lennard-Jones interactions was chosen equal to $1.1$ nm. 
This choice provides reasonable computing efficiency and is sufficient to correctly reproduce 
the properties of interest for all the systems under study.
The van der Waals tail correction terms to the energy and pressure were taken into account.

For each system, a periodic cubic simulation box was set up with $N_\text{wat}=3000$ water molecules 
and  $N_\text{cur}$ curcumin molecules ($N_\text{cur}$ = $2$, $4$, $8$, $12$, $16$ and $20$). 
The initial configuration of particles was prepared by placing first
the $N_\text{cur}$ molecules randomly in the simulation box. Next, $N_\text{wat}$ water molecules
were inserted into the box. 
Each  system  underwent the energy minimization to remove a
possible overlap of atoms in the starting configuration. This was done by applying the steepest
descent algorithm. After the minimization procedure, we performed equilibration of each system with
the time duration $10$~ps at $298.15$~K and $1$~bar using a small time step $0.1$~fs.
The Berendsen thermostat with the time coupling constant of $0.1$~ps and isotropic 
Berendsen barostat with the time constant of $2$~ps were used. 
The compressibility parameter was taken equal to $4.5\cdot 10^{-5}$ bar$^{-1}$ (corresponding to
the bulk water) 
in all simulations. It was observed that
during equilibration, the curcumin molecules start to assembly and form clusters.
At the end of the equilibration run, a few clusters were formed.

After equilibration, the production runs were performed. We used the V-rescale thermostat 
with the time coupling constant of $0.5\ps$ and Parrinello-Rahman barostat with the time 
constant of $2$~ps to perform the production runs. The time step was chosen to be $1\fs$.
The production run was performed in two stages, each of them with time duration of 100 ns. 
At the first stage, we verify if the curcumin molecules form a cluster and 
if it  behaves as a stable entity. At the second stage, a detailed analysis of the 
curcumin cluster in aqueous medium was performed. With such time 
extension of data, we obtain good statistics of events. It is worth mentioning 
that Hazra et al. \cite{Hazra-2014} studied
similar systems by collecting data up to 40~ns only.

\section{Results and discussion}

\subsection{Trends of assembly of curcumin particles in water}

\begin{figure}[!t]
\begin{center}
\includegraphics[width=0.42\textwidth,angle=0,clip=true]{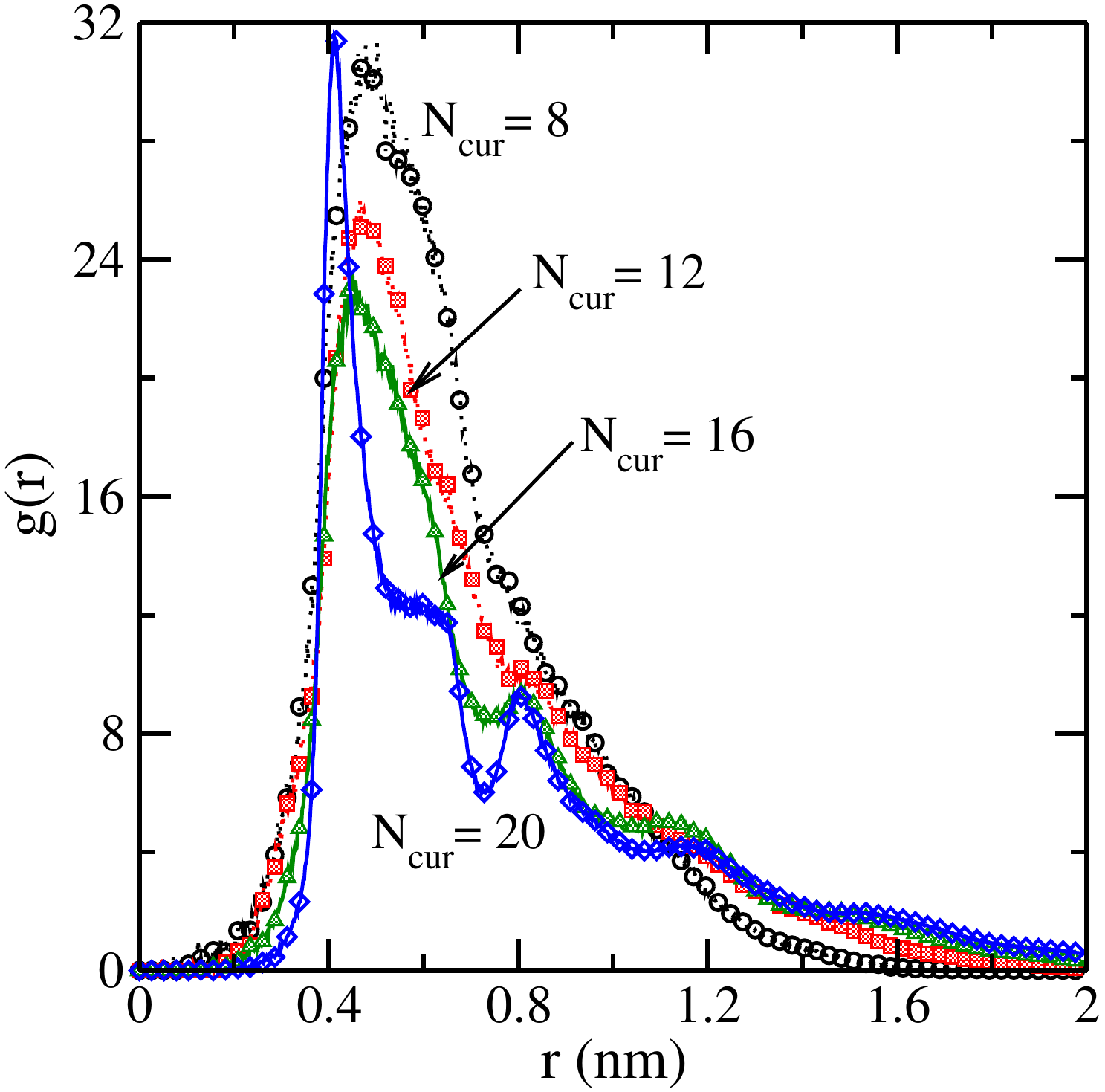}
\caption{(Colour online)\label{fig_rdf_cur} Radial distributions for the center of mass of curcumin molecules
in water ($N_\text{w}=3000$) at room temperature.
}
\end{center}
\end{figure}

To begin with, we 
extracted the time evolution of the center of mass (COM) of curcumin molecules
from the trajectories generated by molecular dynamics for each system under study.
Then, the radial distributions 
for the COM of curcumin molecules were constructed. 
The functions were obtained at the second stage of each production run
and are presented in  figure~\ref{fig_rdf_cur}, for $N_\text{cur}= 8, 12, 16$ and 20.
For $N_\text{cur}= 2$ and 4, the radial distributions are much worse defined because of insufficient
statistics. 
The function $g(r)$ falls to zero at a distance above  $\approx$ 2~nm for all the systems, 
indicating non-homogeneous distribution of curcumin molecules in the box, or in other words showing that 
their aggregation occurred. By contrast, the radial distribution function~(RDF) of water molecules
$g_\text{OW-OW}(r)$ (not shown for economy of space) does not exhibit any peculiarity, 
it behaves as common and  tends to unity at large inter-particle separations.

The function $g(r)$ for curcumin molecules has a single maximum 
in the interval from $r \approx$ 0.4~nm to $r \approx$ 0.5~nm,
if $N_\text{cur}= 8, 12, 16$. The maximum slightly shifts to a smaller distance and
its height decreases with an increasing number of molecules. Moreover, a shoulder 
seen at  $r \approx$ 0.8~nm for $N_\text{cur}= 8$ converts into a local minimum at
a slightly smaller distance for  $N_\text{cur}= 12$ and 16. This behavior reflects the development
of a certain structure upon the assembly of curcumin molecules. The width of the distribution
upon changes from $N_\text{cur}= 8$ to 12 and next to 16 becomes smaller in the interval from 0.3~nm 
up to $r \approx$ 1.1~nm,  whereas the outer part, for $r >$ 1.1~nm, becomes more diffuse
or better say extends to a larger inter-particle separation, if the number of
curcumin molecules increases from 8 to 20. 
Drastic changes of the shape of $g(r)$ occur, if the number
of molecules increases from $N_\text{cur}= 16$ to 20. The first maximum is very narrow
for $N_\text{cur}= 20$
(its height essentially increases upon the change from $N_\text{cur}= 16$ to 20) and the following
wavy structure reflects the ordering of curcumin species within the agglomerate. 

\begin{figure}[!t]
\begin{center}
\includegraphics[width=0.43\textwidth,angle=0,clip=true]{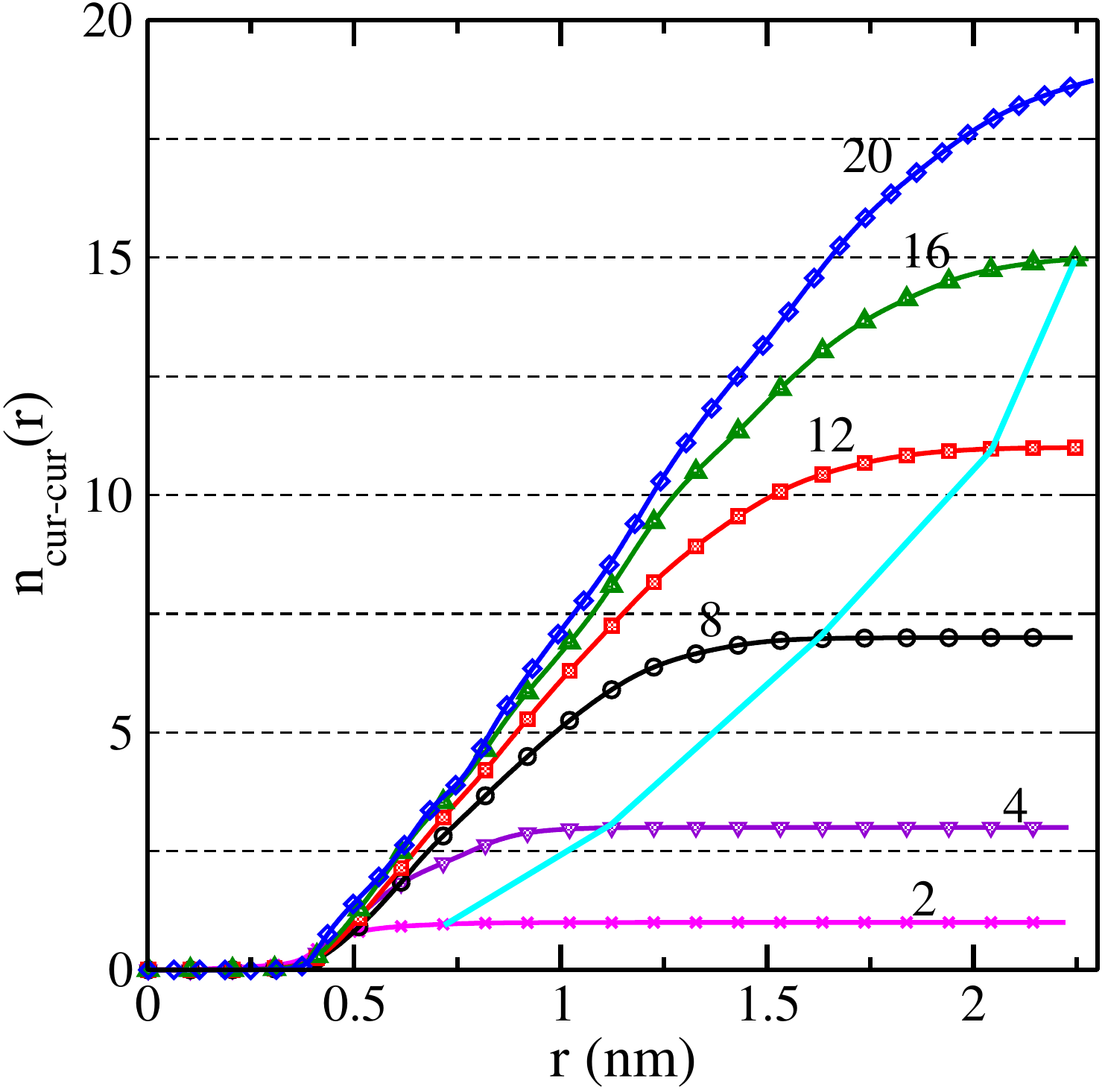}
\caption{(Colour online) \label{fig_nr_cur} Running coordination number of the COM of curcumin molecules
in water ($N_\text{w}=3000$) at room temperature. The number of molecules in each case
is marked in the figure. The cyan line joins the points at which the coordination 
number reaches plateau values with $N_\text{cur}-1$.
}
\end{center}
\end{figure}

An additional insight into the trends of the assembly of curcumin particles in water can be 
obtained from the running coordination numbers defined as follows,
\begin{equation} 
  n_\text{cur-cur}(r) = 4\piup(N_\text{cur}/L^3) \int_0 ^r  g(R)  R^2 \rd R,
\end{equation}
where $L$ is the length of the edge of the simulation box. The corresponding curves are given in 
figure~\ref{fig_nr_cur}. The cyan curve indicates that the size of the agglomerate of
curcumin molecules increases non-monotonously. Apparently, the size tends to the saturation value,
though in order to prove it, one would need to simulate the systems with a much larger number of 
particles than we explore in the present work. 

It is helpful to put the results shown in figure~\ref{fig_rdf_cur} in the context of findings
of other authors. Specifically, Hazra et al.~\cite{Hazra-2014}  provided the radial distribution
of curcumin molecules  within their QC model dispersed in the SPC-E water. The bimodal shape
of the radial distribution  is much stronger pronounced, comparing to our model. 
The wavy structure was also  observed. Unfortunately, neither the number of water molecules
was reported nor the value of the chosen  mole fraction of curcumin. 
Just recall, that our plot concerns the effect of augmenting curcumin concentration in the
solution.

On the other hand, Bonab et al.~\cite{bonab}  used a very similar QC curcumin model 
as Hazra et al. in combination with the SPC water. The authors restricted to the systems with
$N_\text{cur}= 2, 3, 4$ and 5, dispersed in $N_\text{wat} \approx 1500$ water molecules in the box.
The shape of the radial distribution of curcumin 
species differs from ours and from Hazra et al. results, cf. figure~3 
of~\cite{bonab}. Still, the reported dependence of $g(r)$
on concentration is in agreement with our predictions. Namely, the height of the first maximum
decreases with augmenting curcumin concentration (it is located in the interval 
$r \approx 0.4$~nm to $r \approx 0.5$~nm similarly to our data) 
and there exists a distance of crossover of the width
of the distribution (at a distance we report upon the changes from $N_\text{cur}=8$ to 12).

For the sake of better visualization of the structure that the curcumin molecules attain 
in water, we present typical snapshots picked up at the end of the second stage of the
production run for each system. The plots were prepared by using the VMD software~\cite{VMD}. 
The 2D plots in the printed version of the manuscript do not permit to appreciate the entire 
three-dimensional structure, unfortunately. 

\begin{figure}[!t]
\begin{center}
\includegraphics[width=0.4\textwidth,angle=0,clip=true]{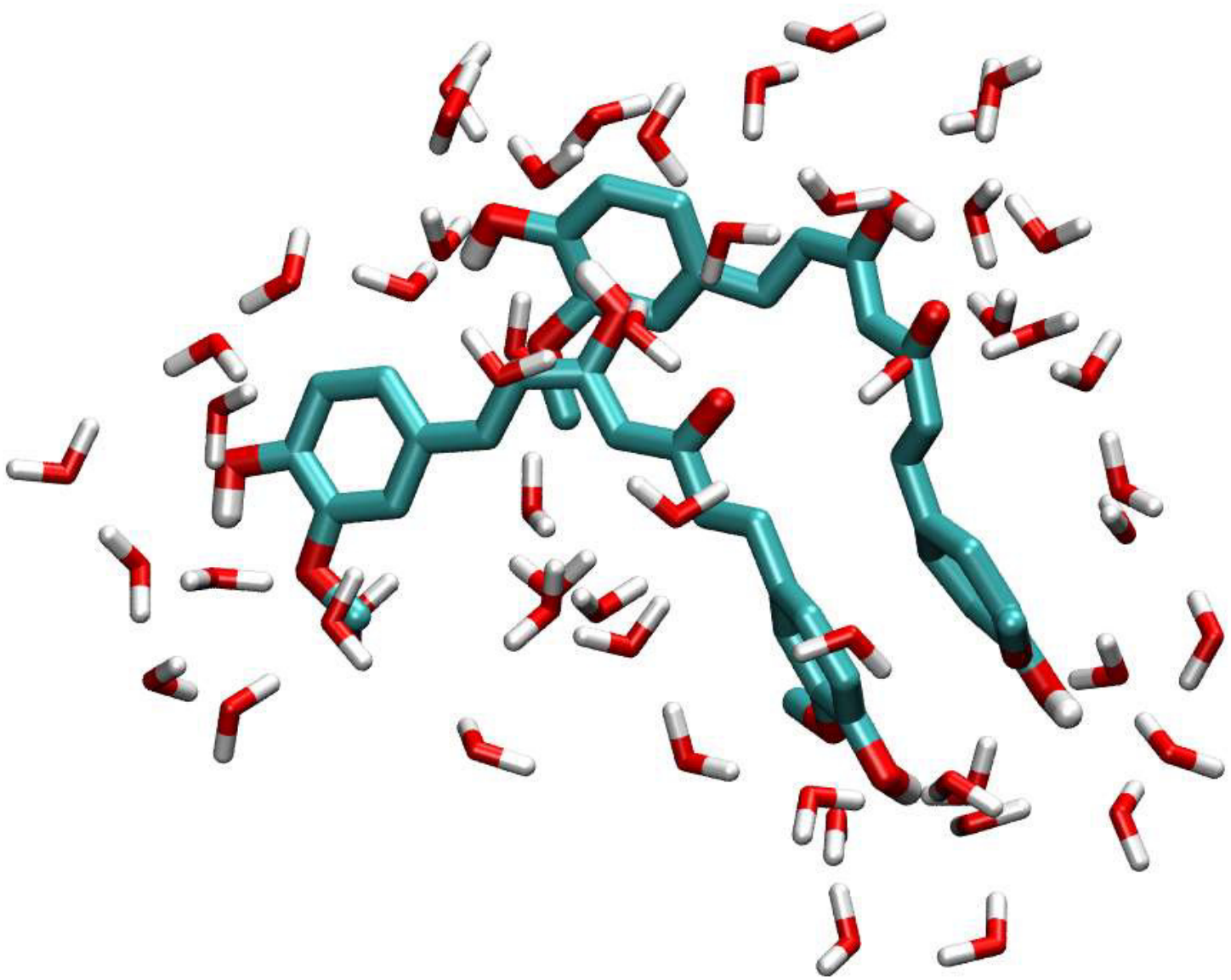} \qquad
\includegraphics[width=0.4\textwidth,angle=0,clip=true]{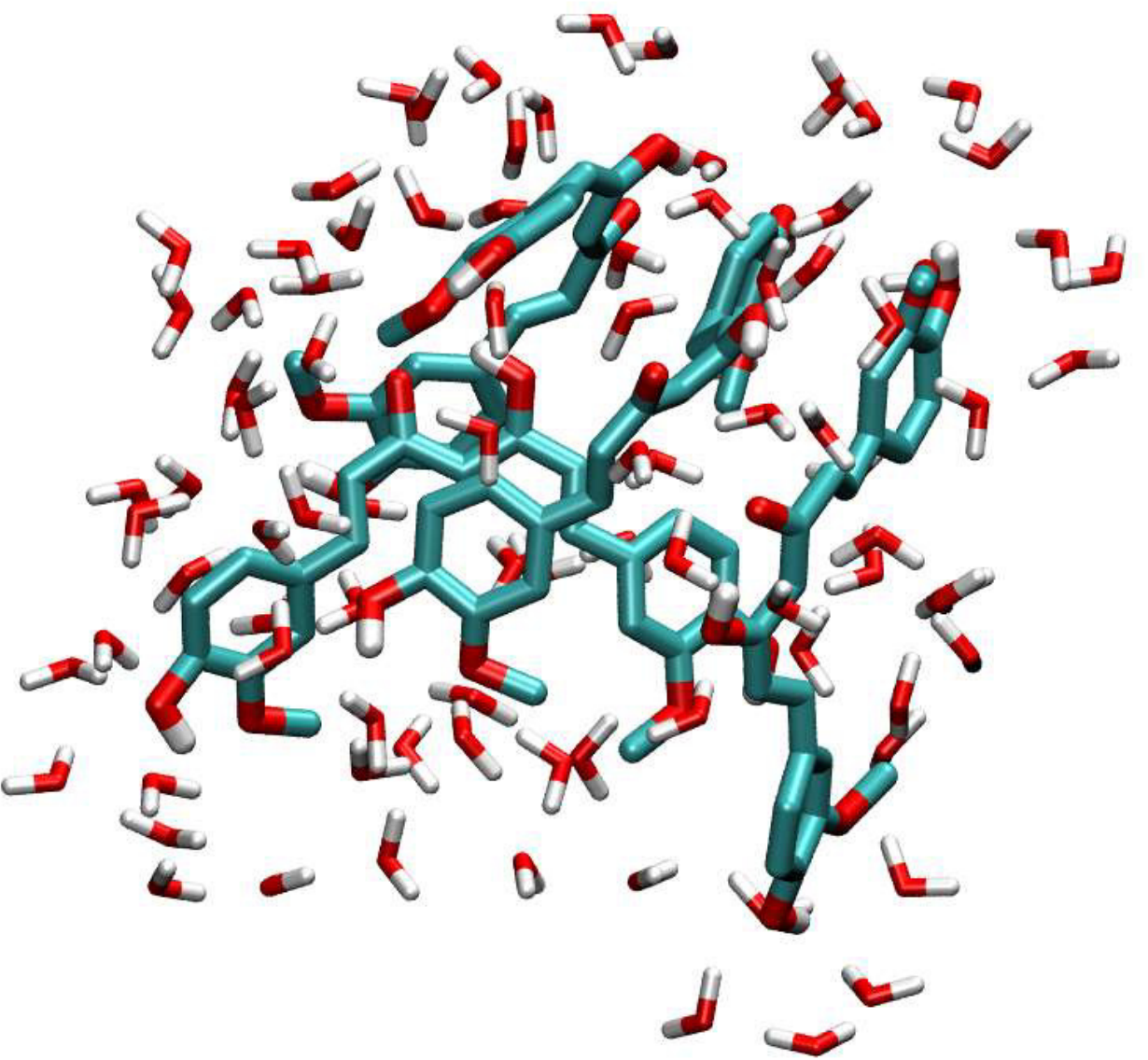}
\caption{(Colour online) \label{fig_clust2} Typical snapshot of the structure attained 
by  2 and 4 curcumin molecules in water, respectively.}
 \end{center}
\end{figure}

Still, in the case of a ``dimer'' ($N_\text{cur}= 2$), we observe a pattern describing trends for 
parallel orientation of molecules, figure~\ref{fig_clust2}. 
Two curcumins are close to each other, water species
hardly penetrate the space between much larger molecules. However, one can notice that there
are water particles approaching the hydrophilic entities of the curcumin molecules.
It is worth mentioning, that the optimized curcumin ``dimer'' structure of the QC
model in the water-ethanol mixture shows a well pronounced parallel motif,
cf. table~1 of~\cite{pereira}.
Much less order apparently exists in the agglomerate of four curcumin molecules 
dispersed in water, figure~\ref{fig_clust2}. 
The parallel motif seen in a dimer seems to be perturbed upon
aggregation of two more curcumin molecules to the dimer.

\begin{figure}[h]
 \begin{center}
 \includegraphics[width=0.4\textwidth,angle=0,clip=true]{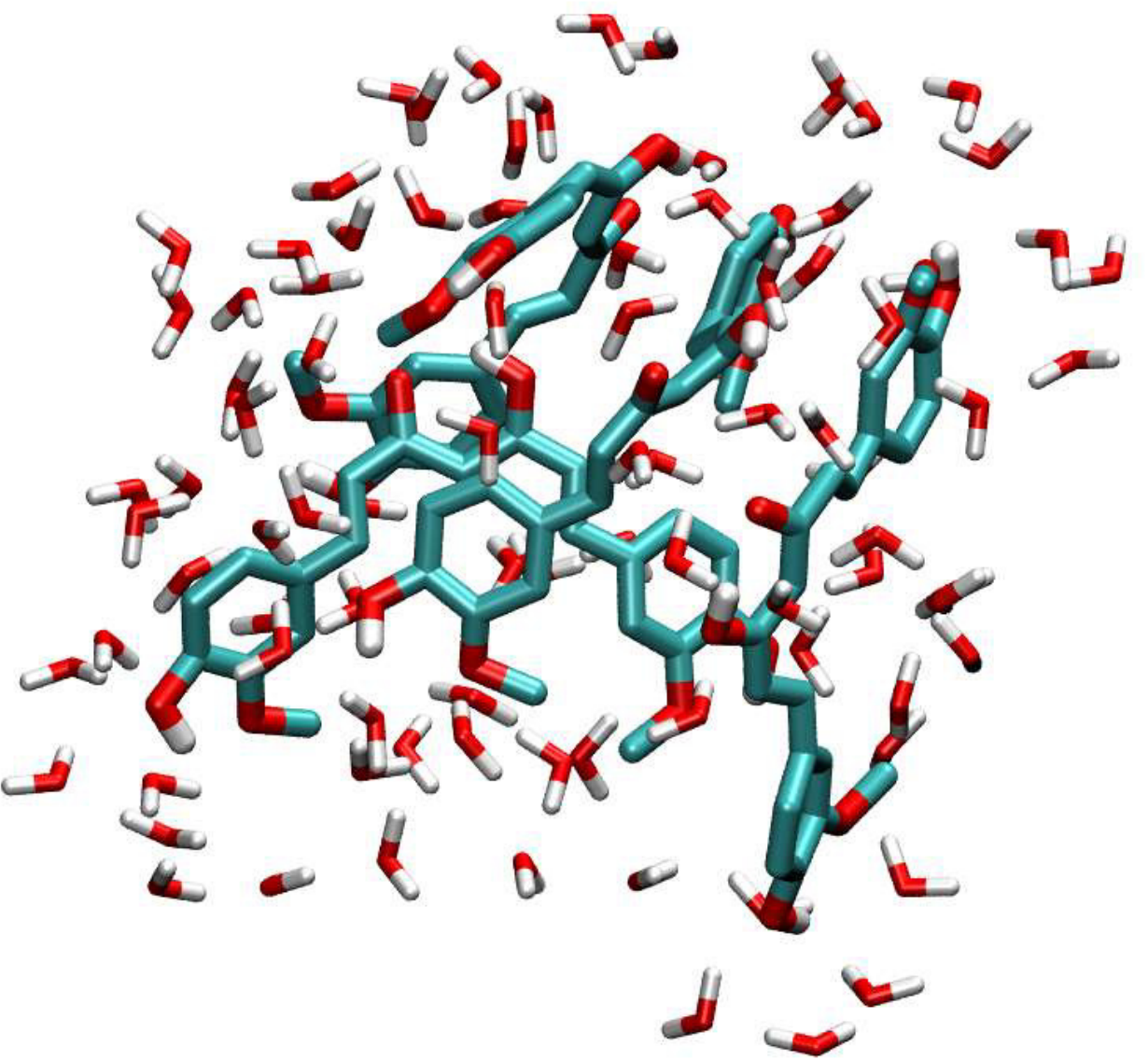} \qquad
 \includegraphics[width=0.4\textwidth,angle=0,clip=true]{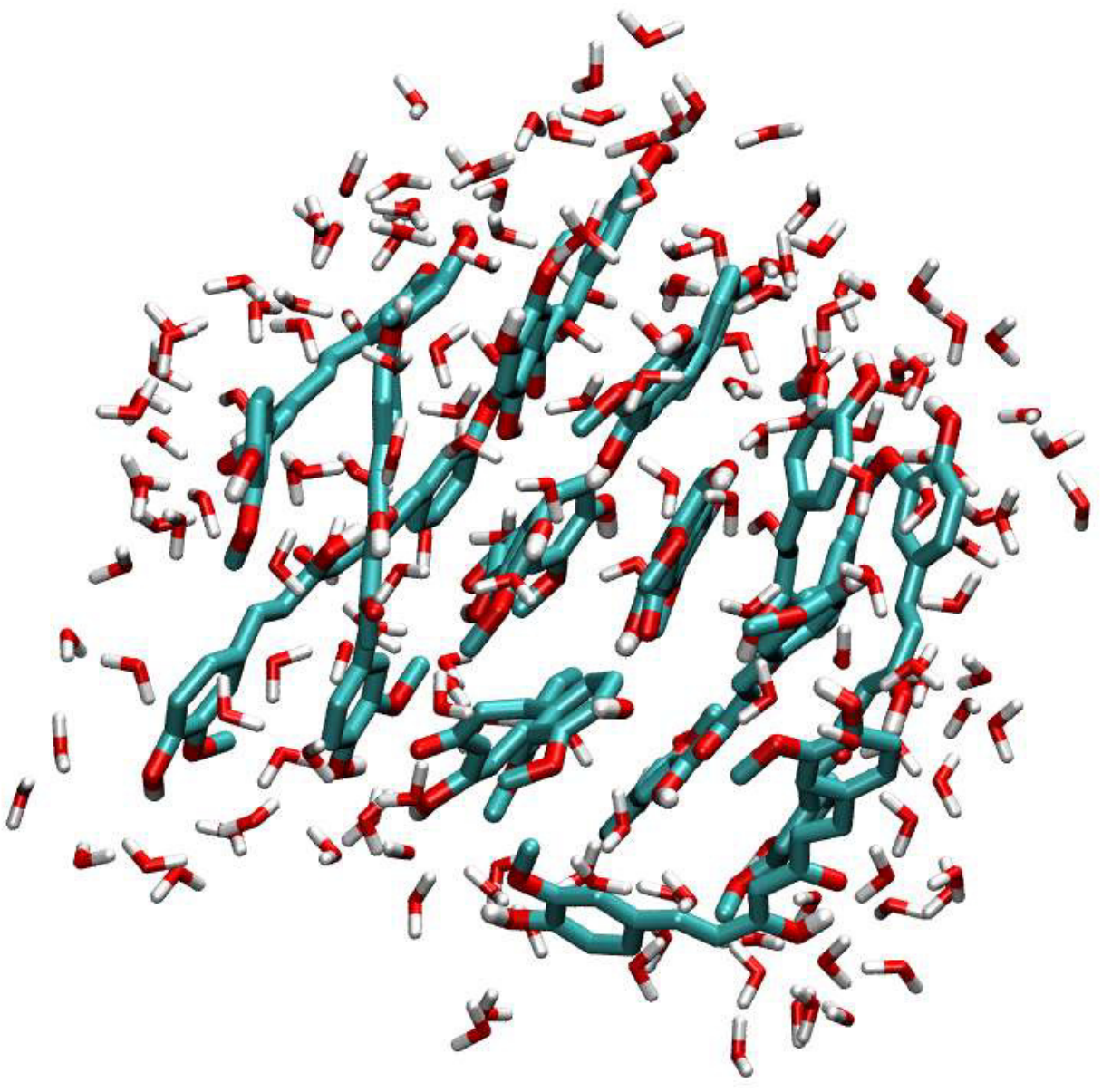}
\caption{\label{fig_clust8}(Colour online) Typical snapshot of the structure 
attained by 8 and 12 curcumin molecules in water.}
 \end{center}
\end{figure}

If the number of curcumin molecules is larger, $N_\text{cur}= 8$ and $N_\text{cur}= 12$,
one can conclude, from both panels of figure~\ref{fig_clust8}, that the trends
for parallel ordering are either diminished (left-hand panel of this figure) or
enhanced and dominate the structure, as in the right-hand panel. In both cases, it seems
that water species are expelled from the interior of agglomerate of curcumin
molecules. This issue will be discussed  below  more in detail.

\begin{figure}[!t]
   \begin{center}
   \includegraphics[width=0.35\textwidth,angle=0,clip=true]{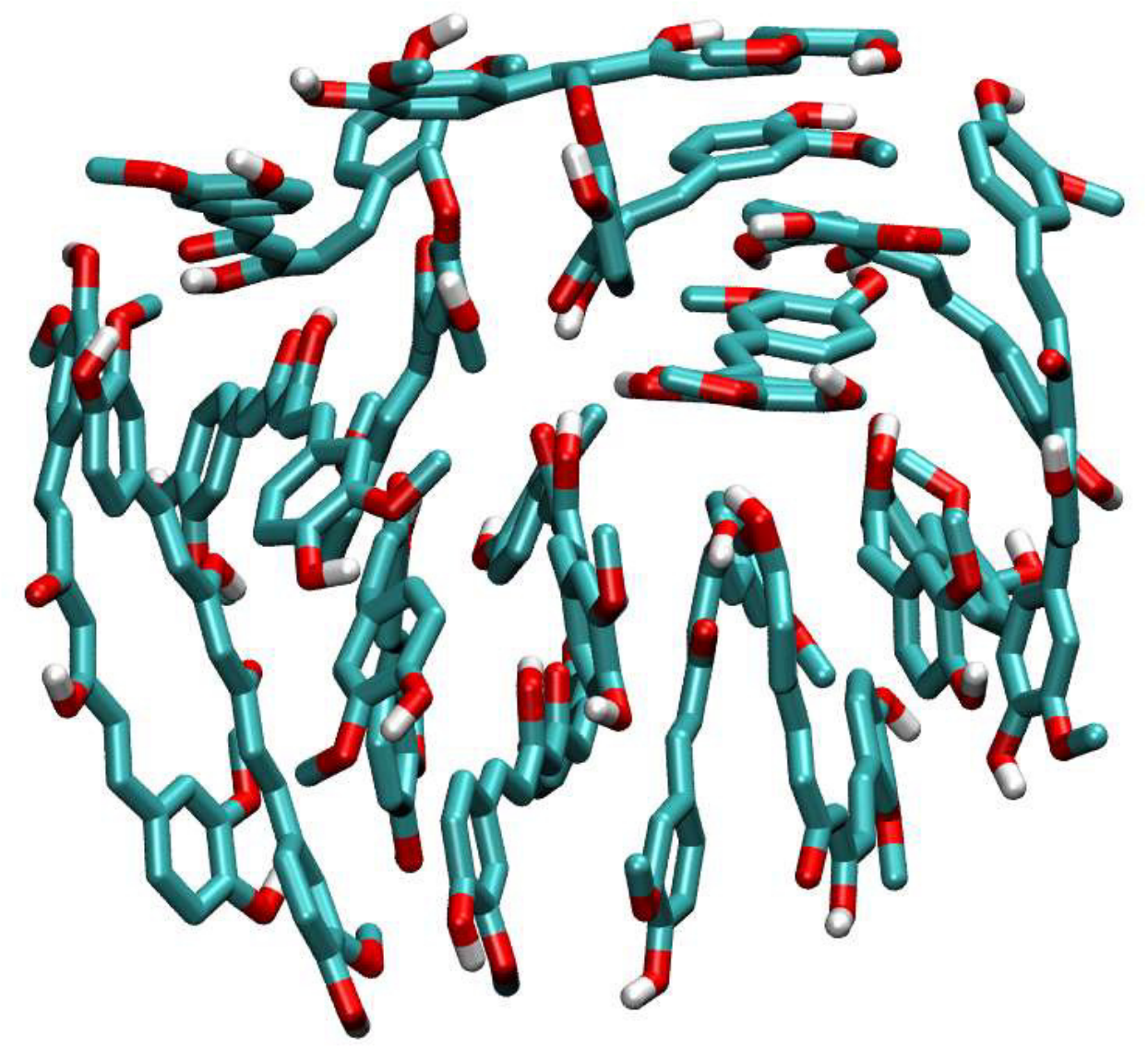} \qquad
   \includegraphics[width=0.35\textwidth,angle=0,clip=true]{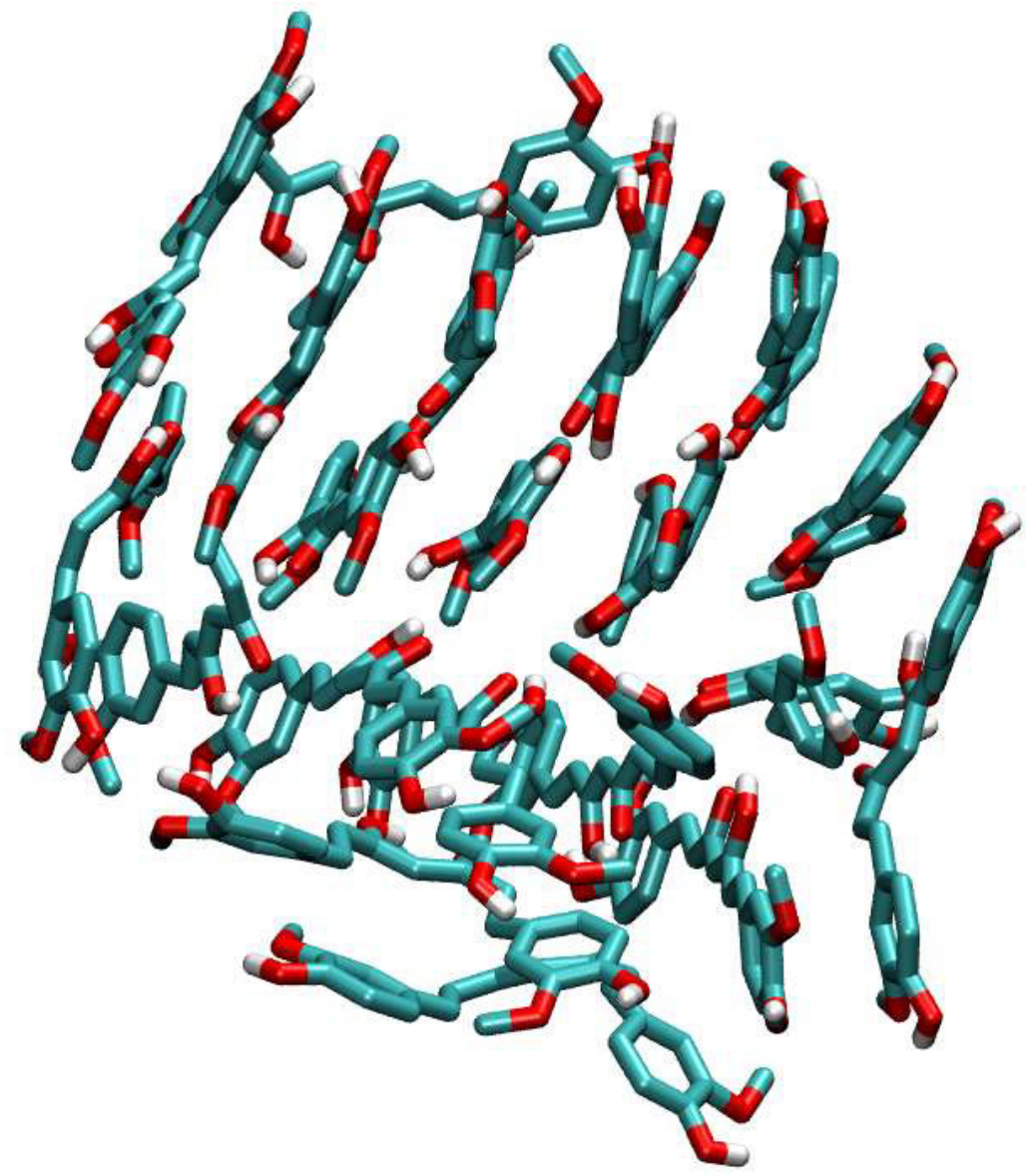}
\caption{\label{fig_clust16}(Colour online) Snapshot of the agglomerates  consisting of 16 and 20 
curcumin molecules in aqueous medium. The water molecules are not shown to
emphasize structural motifs of the distribution of curcumin molecules.}
\end{center}
\end{figure}

Final snapshots in figure~\ref{fig_clust16} concern the systems with a larger number of
curcumin molecules, $N_\text{cur}= 16$ and $N_\text{cur}= 20$,  comparing to previous illustrations.
Water particles are not shown in order to make clearer the distribution of solute 
curcumins. From this visualization, one can get an impression that there exist two
parallel motifs, each of them involving a certain number of molecules. However,
it seems that these two trends compete or, to put it in other words, two
predominant orientations are tilted with respect to each other. We will  
support this hypothesis in qualitative terms below.

The word cluster in the above part of this section  has not been used so far. Now, it is time to do that.
We apply the geometric definition for a cluster, in certain similarity to the geometric definition 
of hydrogen bonds. Namely, taking into account the behavior of the radial distribution
of curcumin molecules in water, figure~\ref{fig_rdf_cur}, we choose the ``cut-off'' distance
$R_{c}=0.75$~nm as a criterion for the cluster definition. 
This distance approximately coincides with the position of the local minimum observed
in the radial distribution of COM of curcumin molecules for the case $N_\text{cur}= 16$ 
and $20$ (figure~\ref{fig_rdf_cur}). Thus, all the molecules satisfying such geometric
criterion along the second stage of the production run are counted as those belonging  
to a cluster. It is worth mentioning that the radial distributions of curcumin
molecules explored with different force field quantitatively differ from the results
of the present model, cf. figure~2 of Hazra et al.~\cite{Hazra-2014}. 
In particular, the second maximum, observed for $N_\text{cur}= 12$, 16 and 20 for the model
in question in figure~\ref{fig_rdf_cur} is small, in contrast to what is reported by 
Hazra et al.~\cite{Hazra-2014} for their QC model. 
However, these authors used the geometric criterion deduced from the local 
minimum of the radial distribution as well. 
In their model $R_{c}=0.84$~nm (for $N_\text{cur}= 8$ and $N_\text{cur}= 12$) 
and $0.916$ for $N_\text{cur}= 16$, see~\cite{Hazra-2014}.
According to our choice of $R_{c}$, majority of curcumin molecules for the present
model will belong to a cluster, as it follows from the $g(r)$ shape.
Only the configurations described by a ``long-range''  tail 
of $g(r)$ would permit the molecules out of clusters. By contrast, for the QC model
of curcumin in water by Hazra et al., the probability to find the molecules farther than
the cluster criterion cut-off is much higher. Therefore, the growth of the clusters
on time is expected to be different in the present OPLS model and for the
QC model from~\cite{Hazra-2014}. Experimental fact is that the curcumin 
in water-rich solutions has
fast aggregation kinetics~\cite{pereira}. However, the experimental time scale and
time evolution in computer simulations are two distinct things.

\begin{figure}[!t]
    \begin{center}
    \includegraphics[width=0.48\textwidth,angle=0,clip=true]{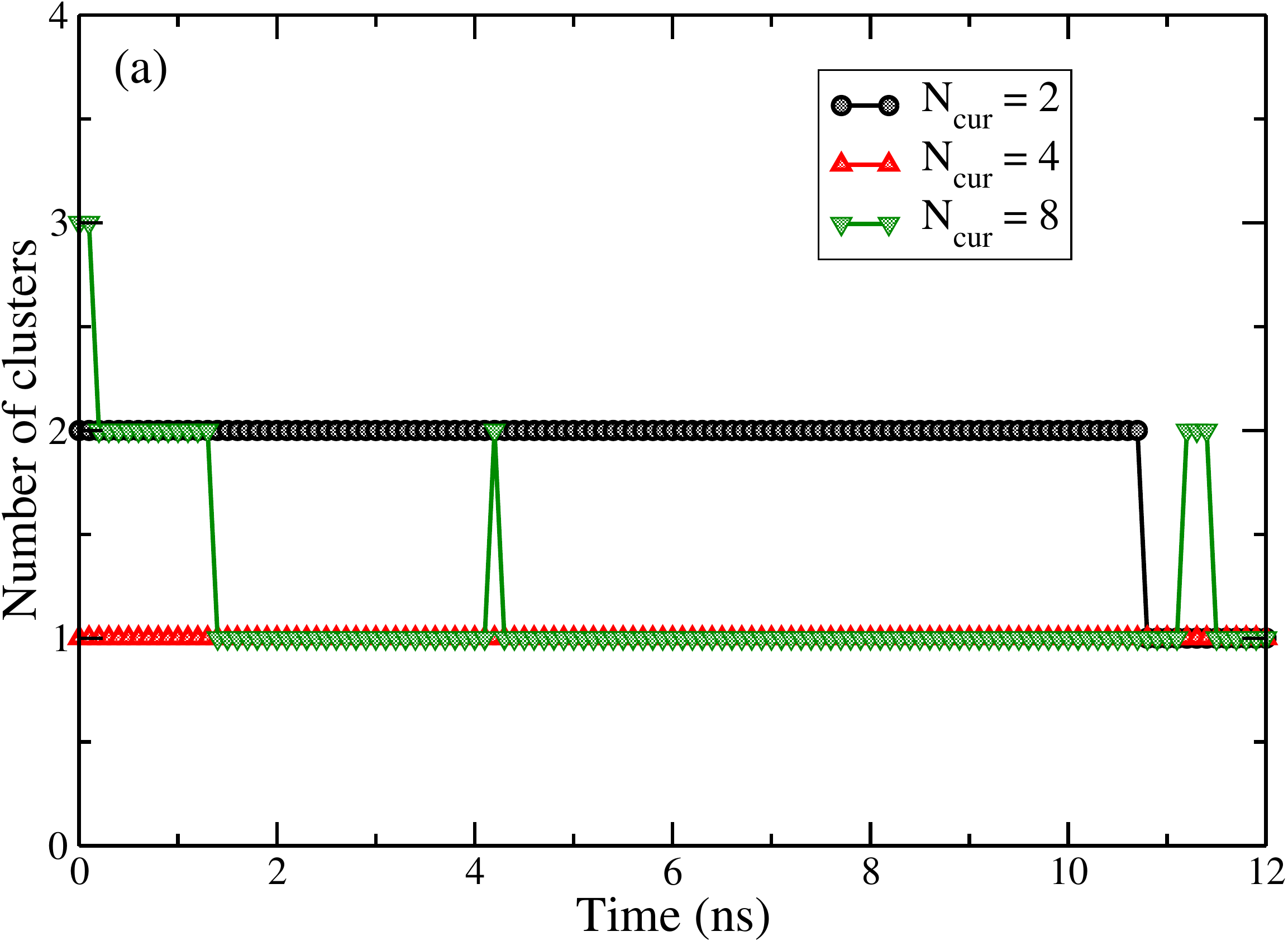}
    \includegraphics[width=0.48\textwidth,angle=0,clip=true]{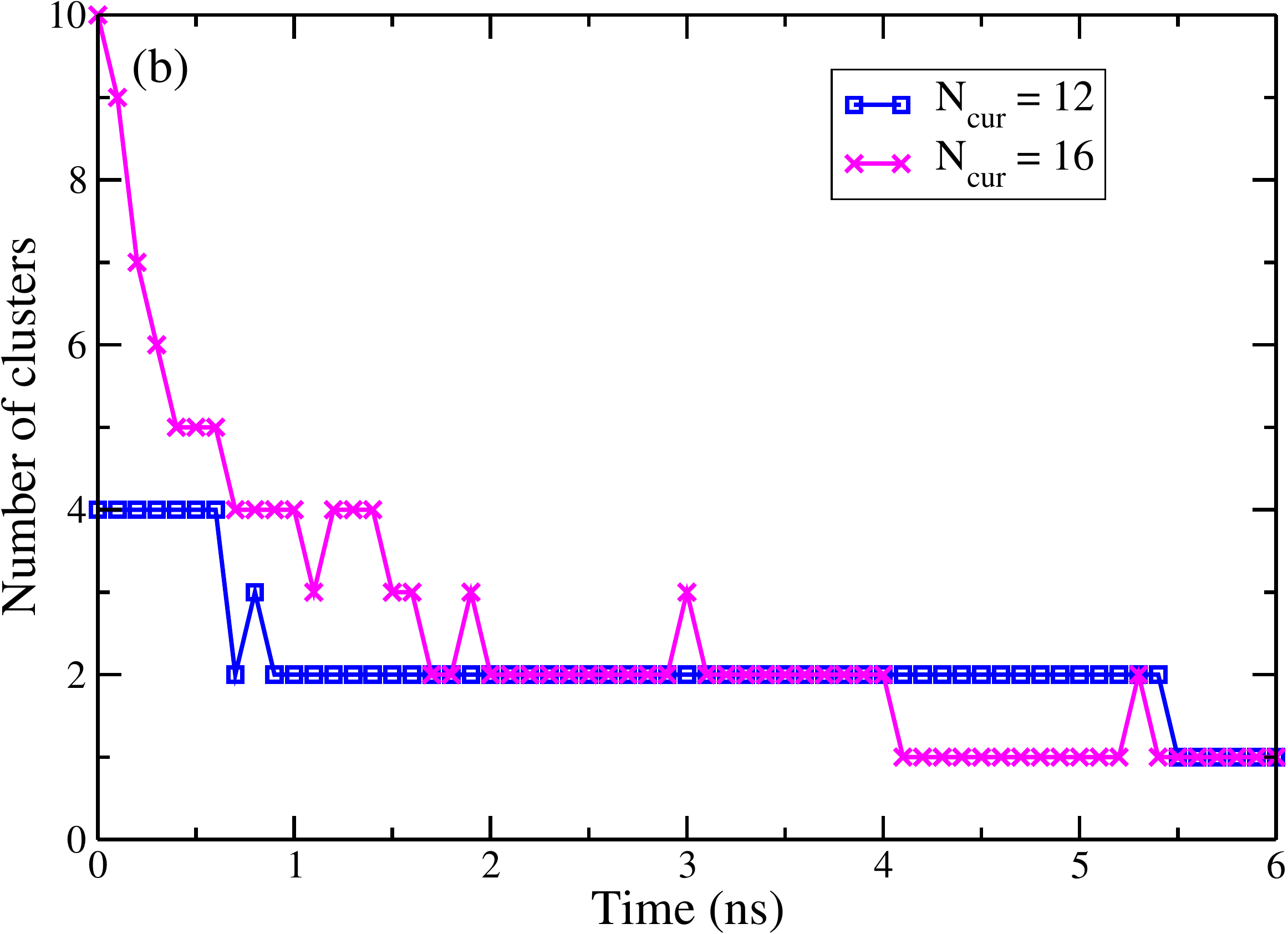}
\caption{\label{fig_clust_size_short}(Colour online) Number of clusters of curcumin molecules at 
the beginning of the first stage of the production run.}
    \end{center}
\end{figure}

\begin{figure}[h]
    \begin{center}
    \includegraphics[width=0.48\textwidth,angle=0,clip=true]{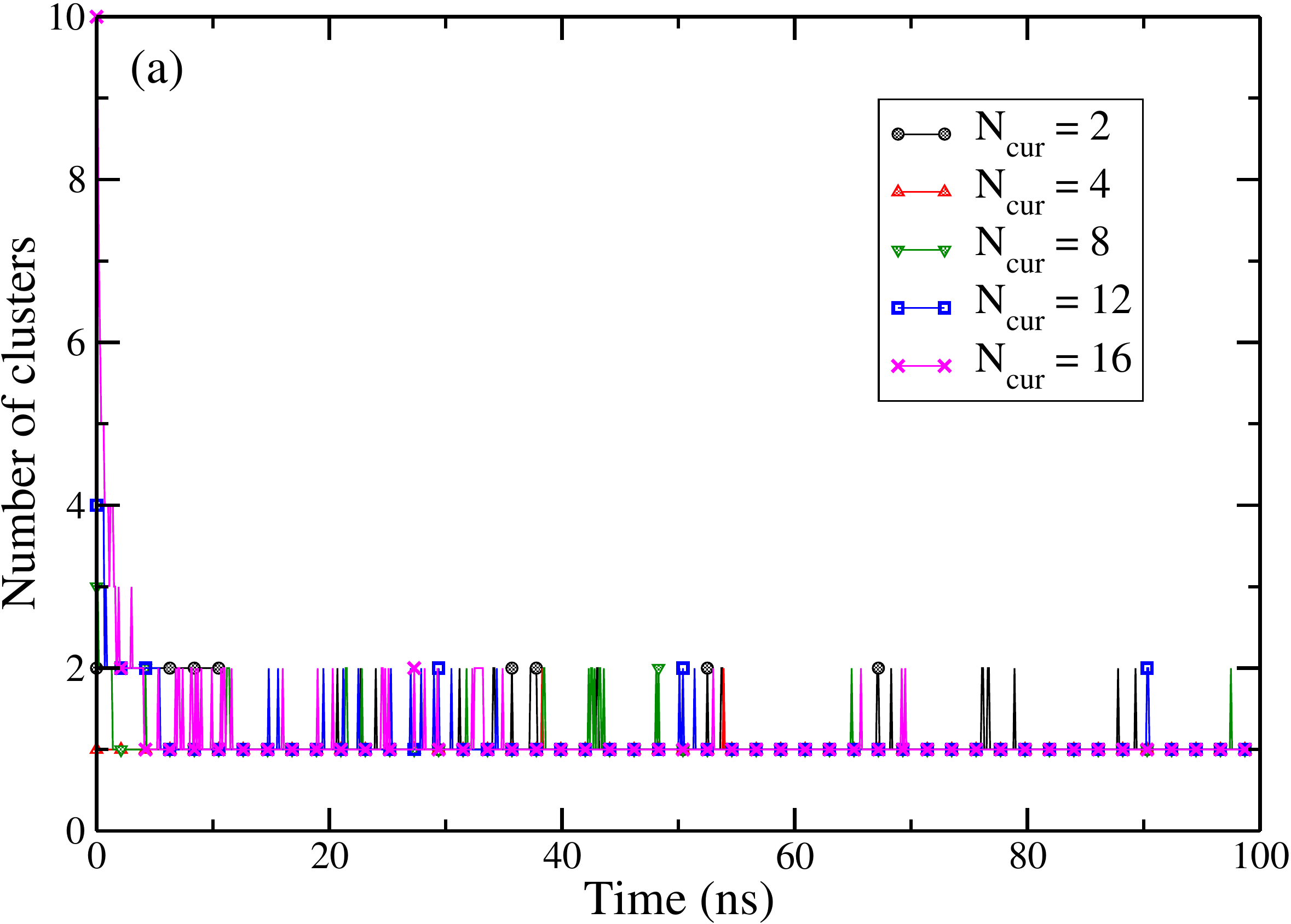}
    \includegraphics[width=0.48\textwidth,angle=0,clip=true]{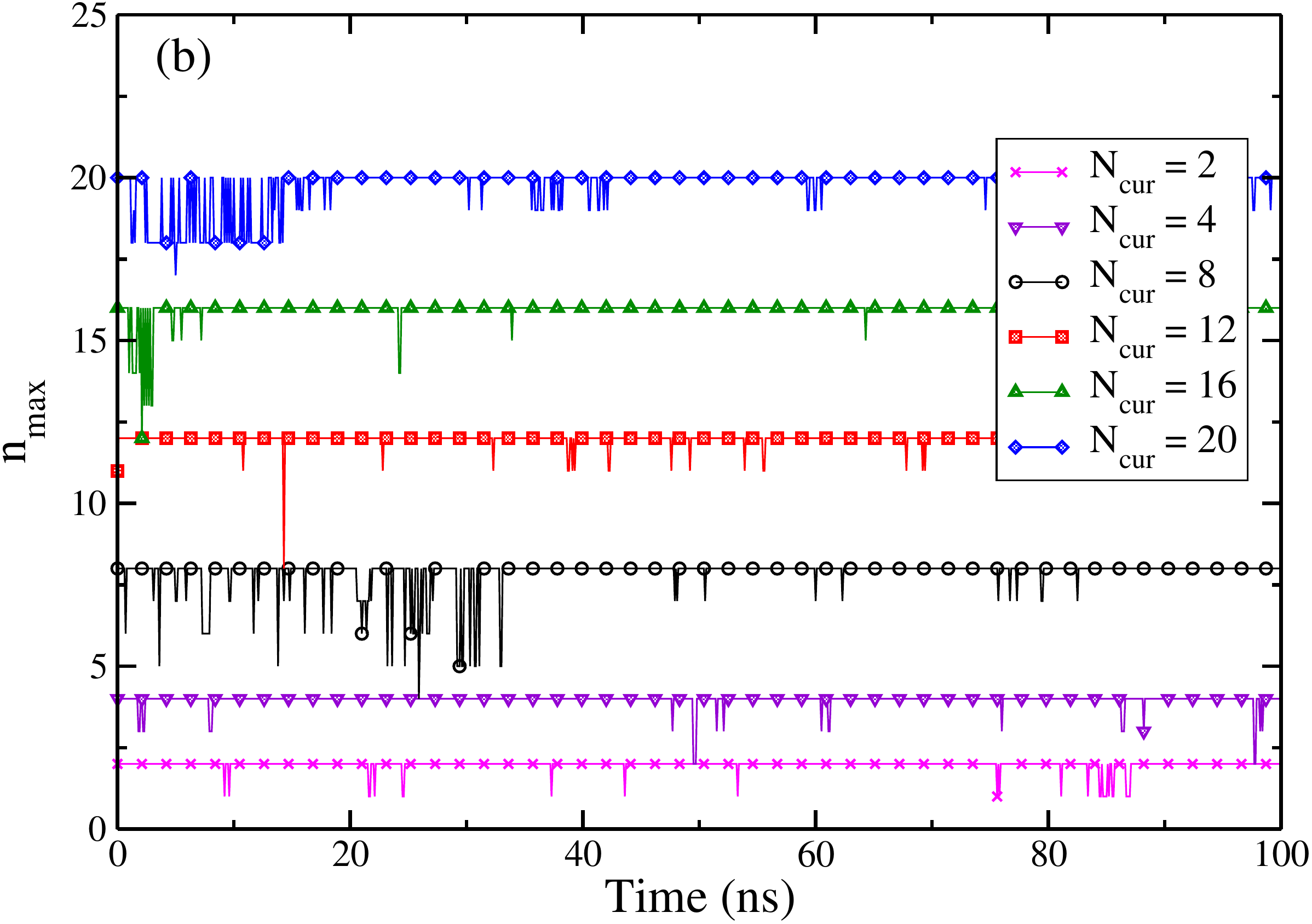}
    \caption{\label{fig_clust_size}(Colour online) Number of clusters of curcumin molecules (panel a) and
the number of molecules in the largest cluster, $n_\text{max}$, dependent on the simulation 
time along the first stage of the production run.}
        \end{center}
\end{figure}

In figure~\ref{fig_clust_size_short} and figure~\ref{fig_clust_size} we present the results
describing the progress of clusters' formation during
a short period of time from the first stage of production run.
It can be seen that the molecules assemble into a  cluster within a few nanoseconds,
figure~\ref{fig_clust_size_short}. However, fluctuations of cluster size
persist
during the entire trajectory corresponding to the first stage
of the production run, figure~\ref{fig_clust_size}.

A detailed description of the formation of cluster for the QC model in water was
given in~\cite{Hazra-2014}. According to figure~\ref{fig_clust2}a of that work, curcumin molecules
do not form a single cluster, even at the end of trajectory of 40~ns. It is difficult 
to prove whether the mechanism of formation of clusters in that work through long-lived
intermediates is universal for curcumin in water or whether it is model specific.

\subsection{On the internal structure of clusters}

In order to analyze the internal structure of clusters formed in each system
under study, we pick up all configurations that yield a ``complete'' cluster, i.e., 
when the number of molecules in the cluster is equal to $N_\text{cur}$
during the entire second stage of the production run with the duration of 100~ns.

\begin{figure}[h]
 \begin{center}
 \includegraphics[width=0.48\textwidth,angle=0,clip=true]{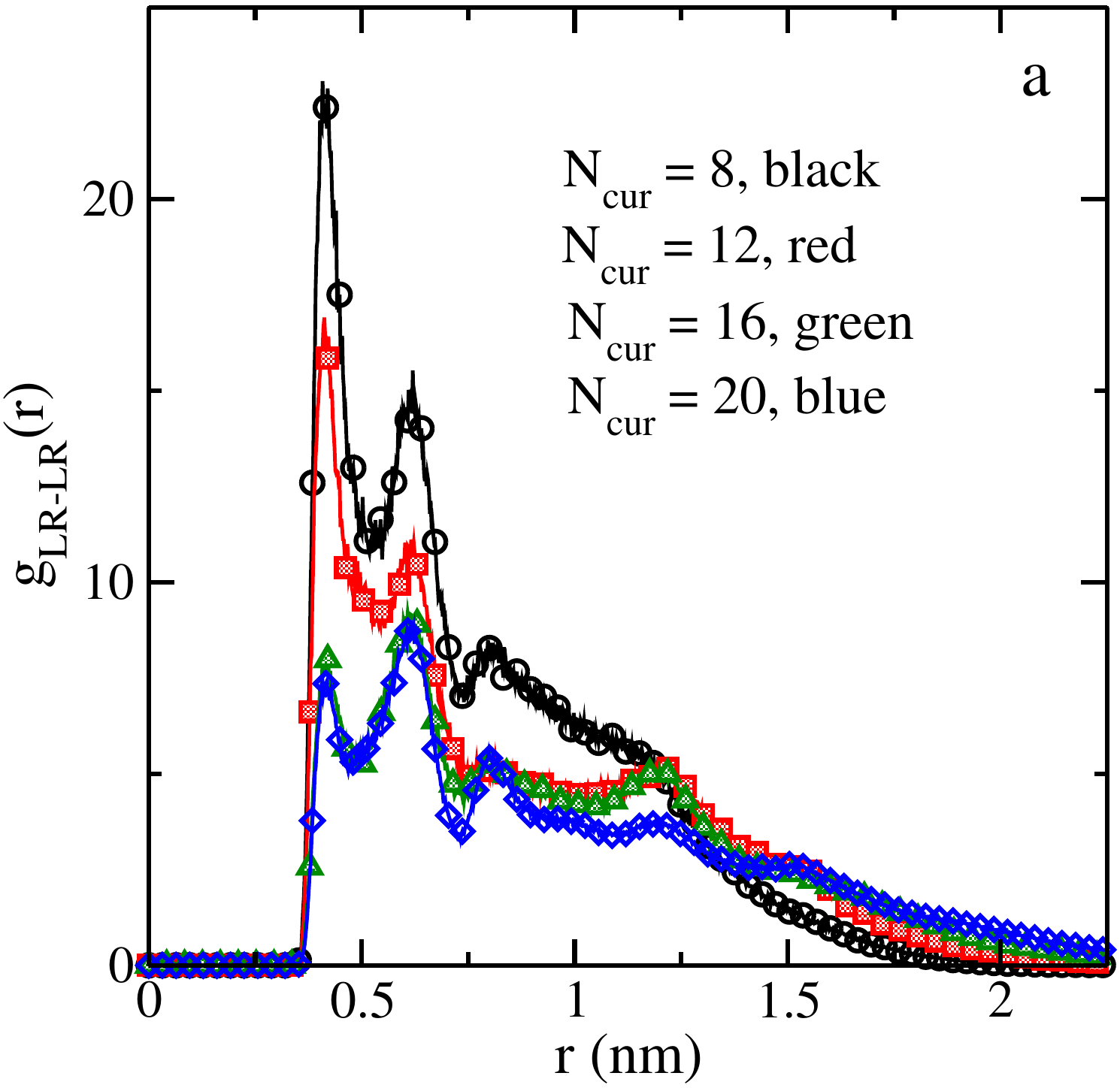} 
 \includegraphics[width=0.48\textwidth,angle=0,clip=true]{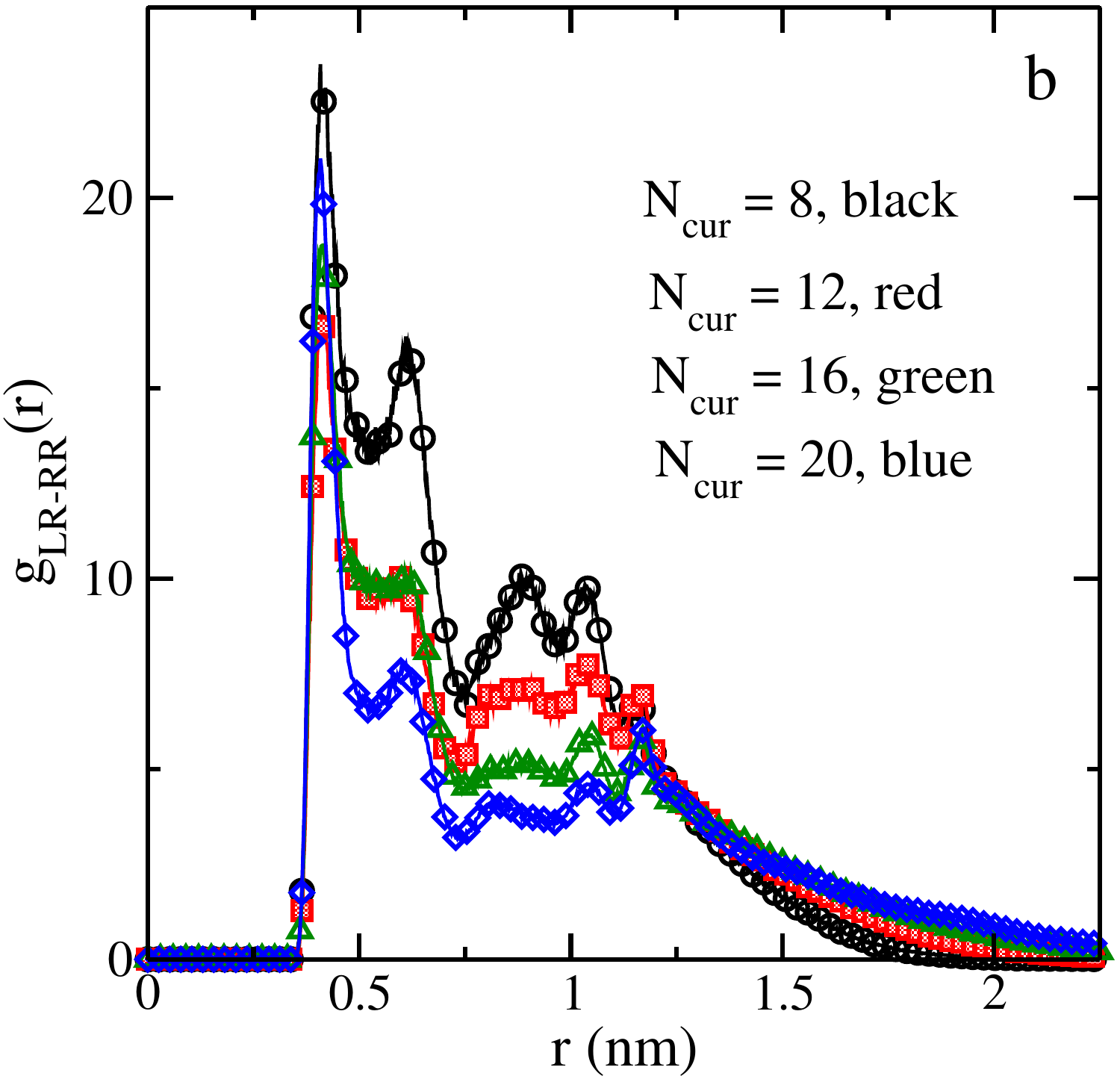}
 \caption{\label{fig_rdf_ring}(Colour online) Pair distribution functions between the 
geometric centers  of the left-hand (LR) and of the right-hand (RR) rings of curcumin molecules: 
LR-LR (panel a) and LR-RR (panel b). In panel~b, the intra-molecular peak is also 
 included.}
 \end{center}
\end{figure}

An overall insight into the radial distribution of curcumin molecules 
with respect to each other in aqueous medium is given above in 
figure~\ref{fig_rdf_cur}. Now, we turn our attention to the distributions
of the structural elements of the curcumin molecules.
Trends of the formation of ordered structures 
can be explored in terms of the radial distributions
between the geometric centers of the left-hand (LR) and of the right-hand (RR) phenyl rings, 
figure~\ref{fig_rdf_ring}. 
In particular, from the  LR-LR RDF, panel a of figure~\ref{fig_rdf_ring},  we see 
that there are three maxima corresponding approximately to the 
distances $0.4$, $0.6$ and $0.8$~nm. The fourth maximum is less pronounced,
at $N_\text{cur} = 8$ it is just a shoulder.
Since the RR-RR counterparts are equivalent to the LR-LR distributions 
up to statistical inaccuracy, they are not presented.

The LR-RR radial distributions, panel b of figure~\ref{fig_rdf_ring},  
are somewhat different comparing to their LR-LR counterparts. 
Nevertheless, they  have two sharp maxima at $0.4$ and $0.6$~nm, the third maximum 
around $0.8$~nm is more disperse, comparing to its LR-LR counterpart.
The fourth maximum is due to intra-molecular structure of the molecule.
These characteristic distances approximately coincide with the distances 
observed in the curcumin-curcumin COM radial distributions in figure~\ref{fig_rdf_cur}.
Thus, the behavior of the LR-LR and LR-RR radial distributions can be interpreted as
describing  parallel and  anti-parallel alignment of phenyl rings of curcumin molecules 
within layer-like structure in the cluster.
 
The height of the first maximum of the LR-LR and LR-RR distributions is approximately
equal if $N_\text{cur}= 8$ and 12. However, in larger clusters ($N_\text{cur}= 16$ and 20),
the first maximum of LR-RR function is much higher,  in comparison to
LR-LR distribution. Thus, in a smaller cluster, parallel and anti-parallel orientations of phenyl rings 
of neighboring molecules are approximately ``equiprobable'' (strictly speaking, 
these functions do not have probabilistic interpretation). 
By contrast, in a larger cluster, anti-parallel
orientation of phenyl rings of nearest molecules dominates over parallel orientation.

\begin{figure}[h]
    \begin{center}
    \includegraphics[width=0.45\textwidth,angle=0,clip=true]{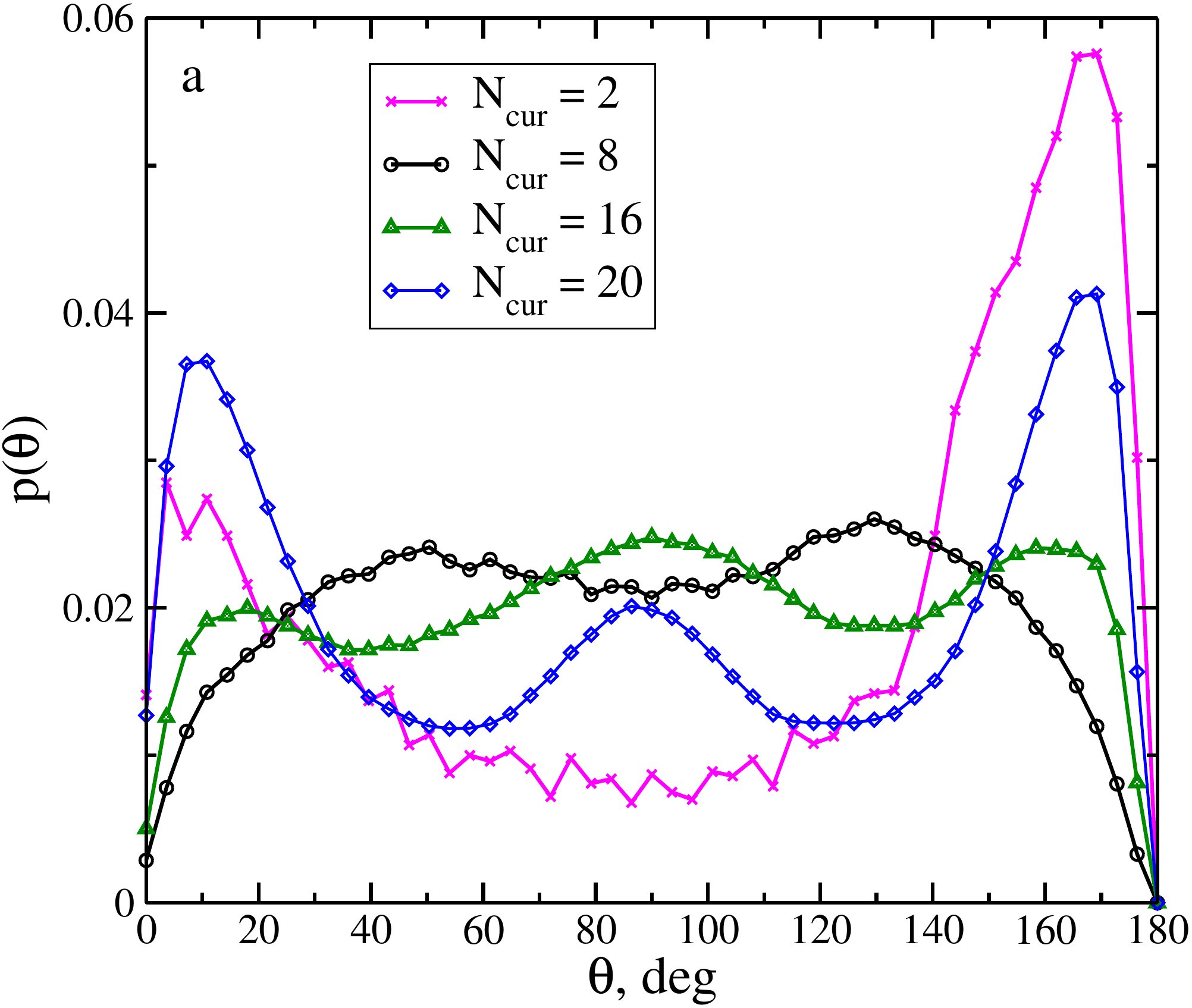} 
    \includegraphics[width=0.45\textwidth,angle=0,clip=true]{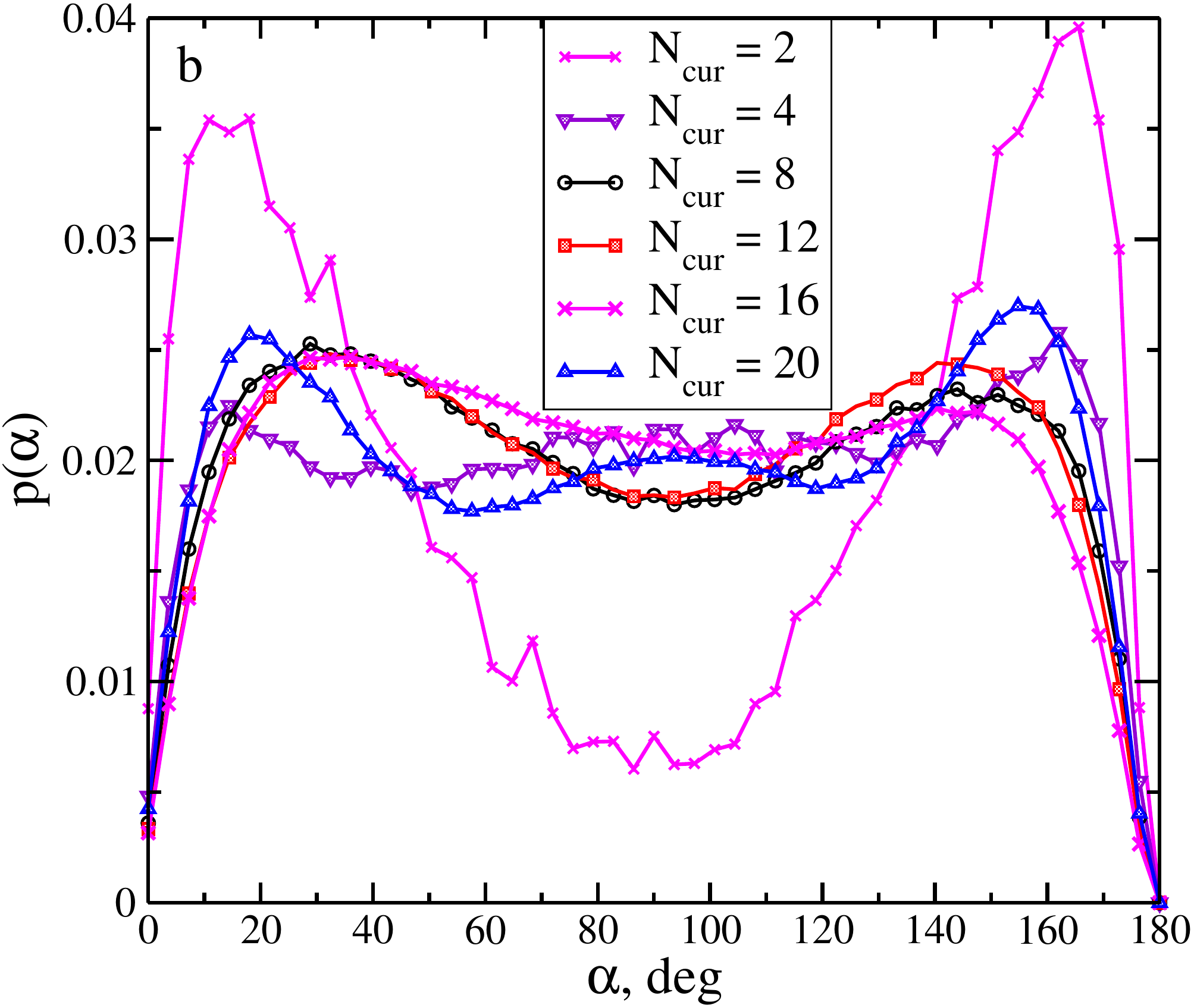}
 \caption{\label{fig_ang_corr}(Colour online) Probability density distribution of the angle between the axes 
of two curcumin molecules in a cluster of size $N_\text{cur}$ (panel a).
Probability density distribution of the angle between vectors normal to a ring of two curcumin molecules (panel b).
                }
        \end{center}
\end{figure}

Trends for alignment of curcumin molecules were explored  more in detail 
by calculating the angular probability distribution functions 
$p(\theta)$ for angles $\theta$  between axes of a pair of curcumin molecules 
(figure~\ref{fig_ang_corr}a).
The axis of a molecule  is defined as a vector between the geometry center of 
the left-hand (LR) and the right-hand (RR) phenyl rings.
The distribution $p(\theta)$ describes  an axial alignment of the 
curcumin molecules  with respect to each other.
From the maxima of $p(\theta)$ in figure~\ref{fig_ang_corr}a, one can conclude that curcumin 
molecules can be oriented 
not only parallel (or anti-parallel), but also normal to each other. 
However, the nearest neighbors tend to solely parallel (anti-parallel) ordering,
and the perpendicular orientations may correspond to molecules from different domains 
formed within the cluster, or it can appear due to the perturbations  
by the proximity of curcumin-water interface.

Another angular probability distribution function $p(\alpha)$ was calculated to analyze the
planar alignment of curcumin molecules, 
where $\alpha$ is an angle between vectors normal to phenyl ring planes 
(figure~\ref{fig_ang_corr}b). In this case, one can also observe a 
preferable orientation of the molecular planes, that are either nearly parallel or 
perpendicular to each other. 
This behaviour indicates that curcumin molecules tend to form a layered structure in a cluster, 
although it can be quite distorted due to a small size of clusters.

The parallel and anti-parallel alignment of curcumin molecules in a cluster 
was discussed by Hazra et al. in \cite{Hazra-2014}. 
However, the perpendicular 
orientations of curcumin molecules in the cluster have not been reported 
since they restricted to the angular probability distribution to the nearest neighbors only.

Trends for axial orientational order of curcumin molecules in a cluster can be
captured by using other mathematical descriptors. Namely, 
to inspect this phenomenon, the order parameter, $S$, of molecules in clusters 
of sizes $N_\text{cur}=12$, $16$ and $20$ was calculated
from the order parameter tensor,
\begin{equation}
\hat{Q}_{\alpha\beta}=\frac{1}{2 N_\text{cur}}\sum_{i=1}^{N_\text{cur}}
\left(3u_{i,\alpha}u_{i,\beta}-\delta_{\alpha\beta}\right), 
\qquad
\alpha,\beta\in \lbrace x,y,z\rbrace,
\label{tensor}
\end{equation}
where $\mathbf{u}_{i}$ is the vector describing the orientation of a molecule $i$,
and  $\delta$ is the Kroneker symbol. 
From the diagonalization of tensor $\hat{Q}$, the uni-axial order 
parameter $S=\frac{3}{2}\lambda_{+}$ can be obtained. Here,
 $\lambda_{+}$ is the largest eigenvalue, and the corresponding eigenvector 
is the director.

\begin{figure}[h]
   \begin{center}
   \includegraphics[width=0.4\textwidth,angle=0,clip=true]{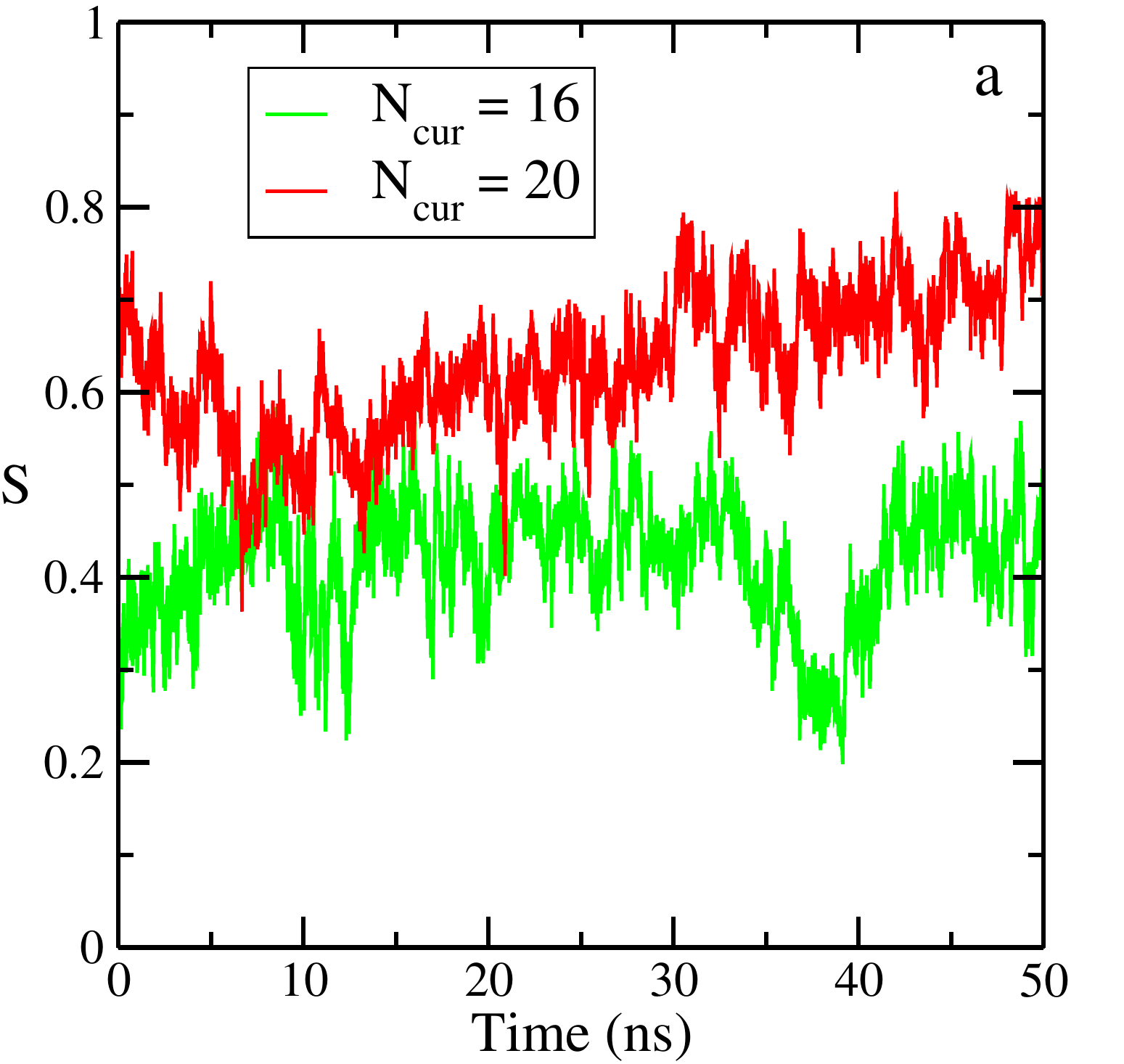}
   \includegraphics[width=0.4\textwidth,angle=0,clip=true]{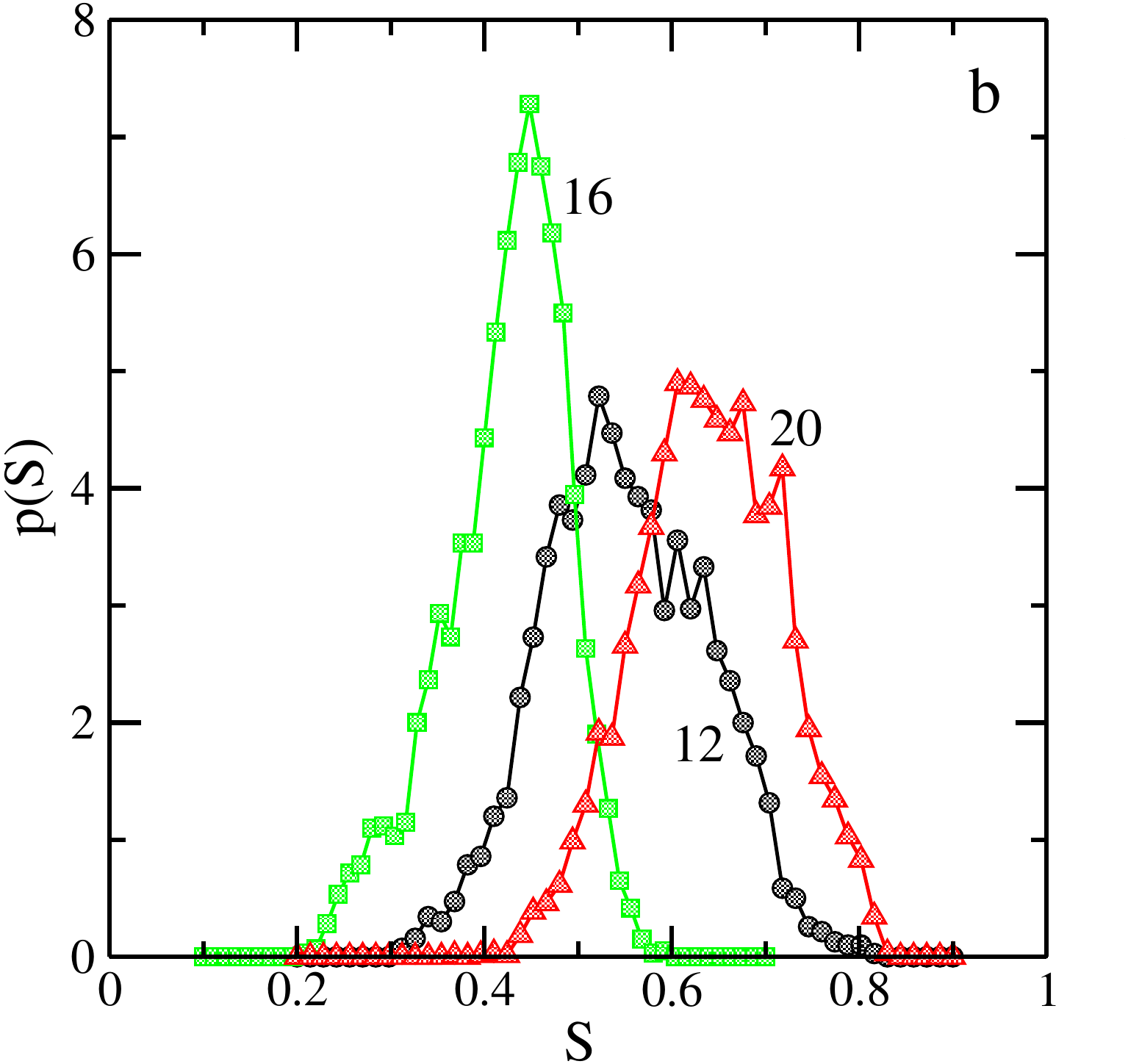}
   \caption{\label{fig_S2}(Colour online)  Order parameter of $N_\text{cur}$ curcumin molecules in a cluster dependent on time~(a).
Probability density distribution of order parameter of $N_\text{cur}$ curcumin molecules in a cluster (b).
}
 \end{center}
\end{figure}

The vector $\mathbf{u}_{i}$, to characterize the orientation, is taken along the line 
connecting geometric centers of the left-hand (LR) and the right-hand (RR) phenyl rings 
of each of curcumin molecules.
Then, the order parameter, $S$, is calculated at each $10$~ps during last $50$~ns of 
the second stage of the production run. This is shown in figure~\ref{fig_S2}a 
(case $N_\text{cur}=12$ is not presented in order to avoid overload of this panel of the figure). 
With these data available, the  probability density distribution of the order 
parameter, $p(S)$, is obtained. 
From figure~\ref{fig_S2}b one 
can see that molecules in the cluster with $N_\text{cur}=20$ are well uni-axially ordered,
the average order parameter is equal to $S\approx0.65$. The clusters with a smaller 
number of curcumin molecules $N_\text{cur}=16$ and $12$ are characterized by a 
less pronounced uniaxial motif.
Moreover, the average order parameter for $N_\text{cur}=12$ appears to be 
higher than for $N_\text{cur}=16$, $S\approx0.5$ and $S\approx0.4$, respectively.

This behavior can be traced by recalling the non-monotonous changes of the 
height of the first maximum for curcumin-curcumin RDF in figure~\ref{fig_rdf_cur}
with the increasing number of curcumin molecules. 
Namely, this function indicates the appearance of an additional characteristic length
when $N_\text{cur}$ increases from 12 to 16. In addition, these trends 
can be elucidated from the snapshots described above.
In summary, the cluster growth is not necessarily accompanied by a monotonous growth
of the uniaxial order parameter.

\begin{figure}[!t]
  \begin{center}
   \includegraphics[width=0.4\textwidth,angle=0,clip=true]{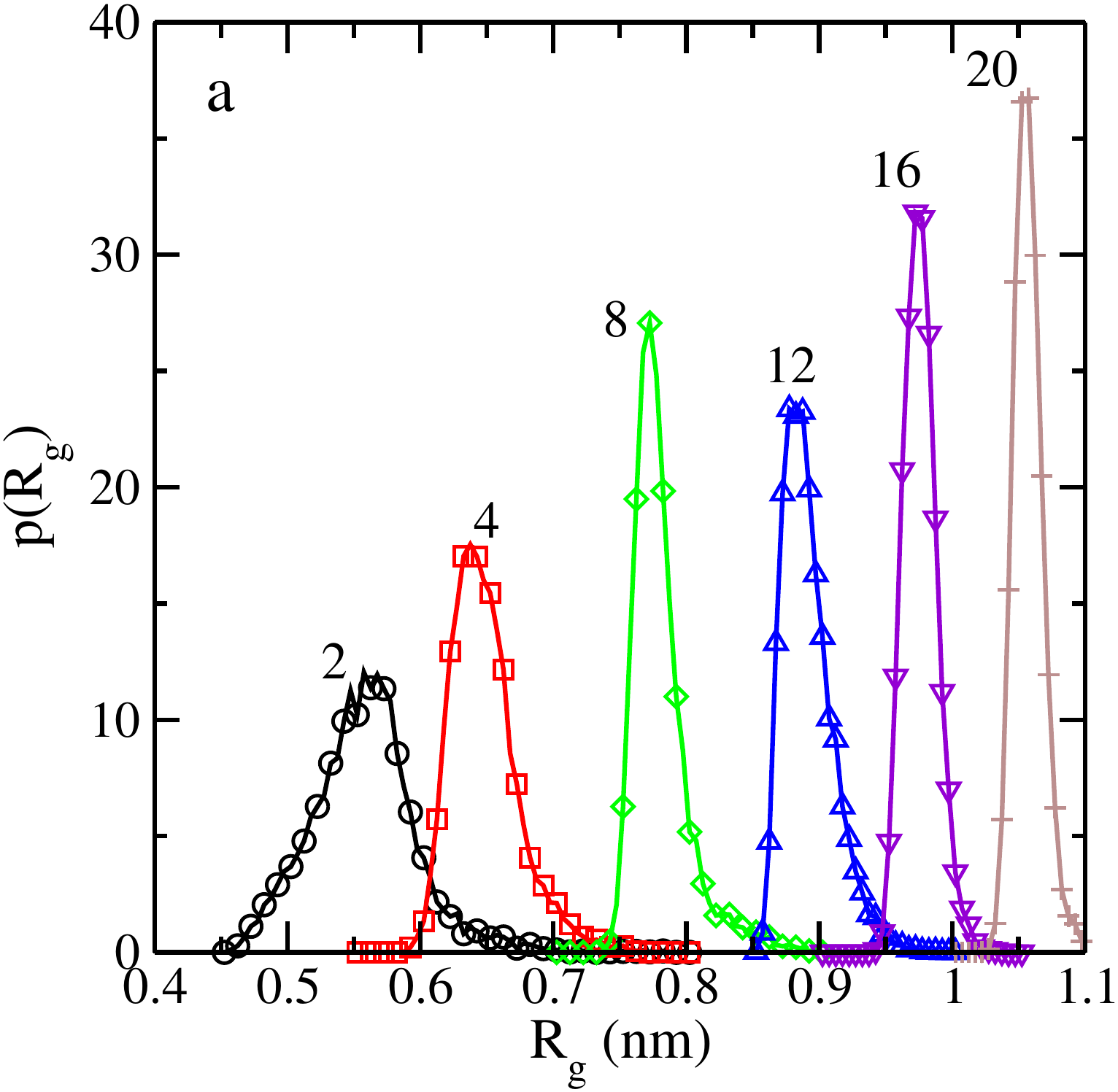}
   \includegraphics[width=0.4\textwidth,angle=0,clip=true]{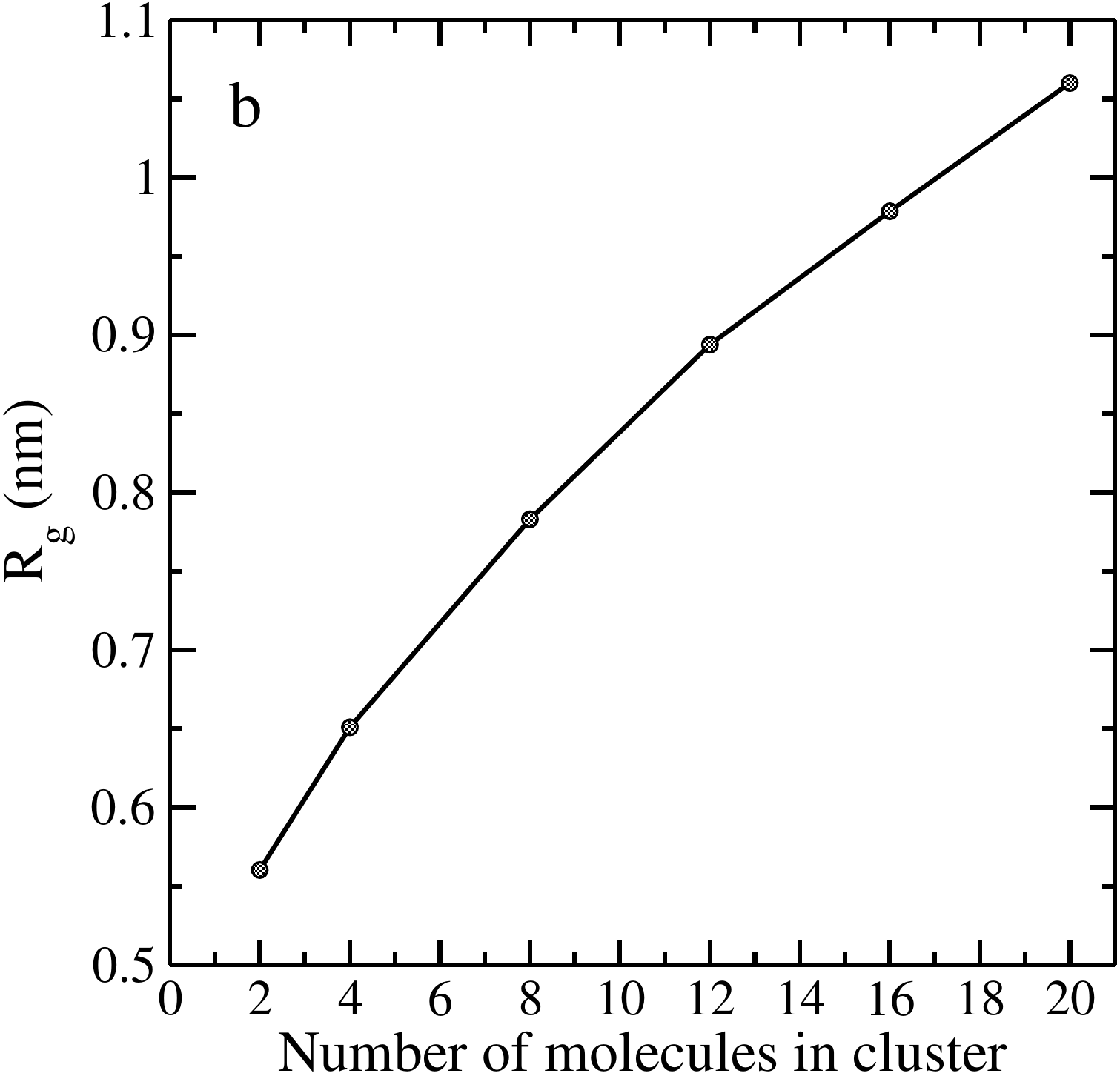}
 \caption{\label{fig_hist_gyr}(Colour online) Probability density distribution of $R_\text{g}$ of the largest cluster 
of curcumin molecules (left-hand panel). Radius of gyration, $R_\text{g}$, as a function of the number 
of curcumin molecules $N_\text{cur}$ (right-hand panel).
 }
\end{center}
\end{figure}

Since we consider the clusters consisting of a small number of particles $N_\text{cur}\leqslant20$, 
they are not large in space.	 
For example, in the case of $N_\text{cur}=20$, our estimate from the corresponding 
radial distribution  yields a size about $2.2$~nm. 
To get a deeper insight into clusters size change, 
we  calculated the probability distribution functions for the radius 
of gyration of clusters, figure~\ref{fig_hist_gyr}a. The gyration radius is a good 
descriptor for the cluster compactness.

We observe that the probability density distribution 
of the gyration radius is unimodal and possesses a certain dispersion. Dispersion
of the distribution  decreases upon increasing $N_\text{cur}$.
The dependence of the average radius of gyration on the number of curcumin molecules in a 
cluster, $N_\text{cur}$, is shown in figure~\ref{fig_hist_gyr}b.
It indicates  a monotonous growth of the geometrical size of a cluster 
with $N_\text{cur}$ from $R_{g}=0.56$~nm up to $1.06$~nm.
Another estimate for the size of curcumin cluster follows from the radial 
distribution of water molecules with respect to the COM of the cluster. 
It is discussed below.

\subsection{On the self-diffusion coefficient of curcumin species}

Obviously, the dynamic properties of curcumin cluster are affected by its size.
One of the methods to obtain the self-diffusion coefficients is from the
mean square displacement of species. The mean-square displacement (MSD) is
obtained from the simulation trajectories during the second stage of production runs. 

\begin{figure}[!t]
\begin{center}
\includegraphics[width=0.45\textwidth,angle=0,clip=true]{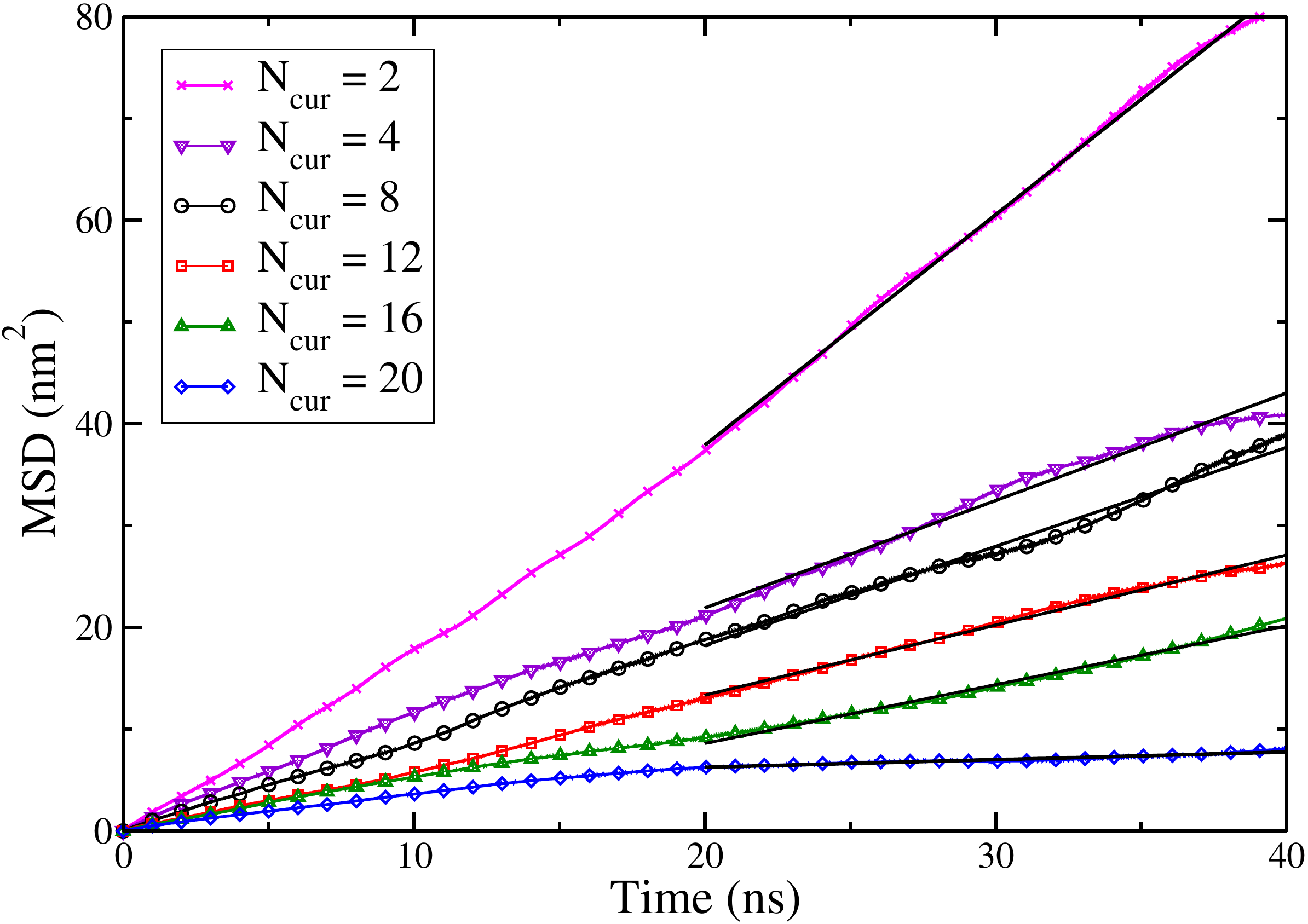}
\includegraphics[width=0.45\textwidth,angle=0,clip=true]{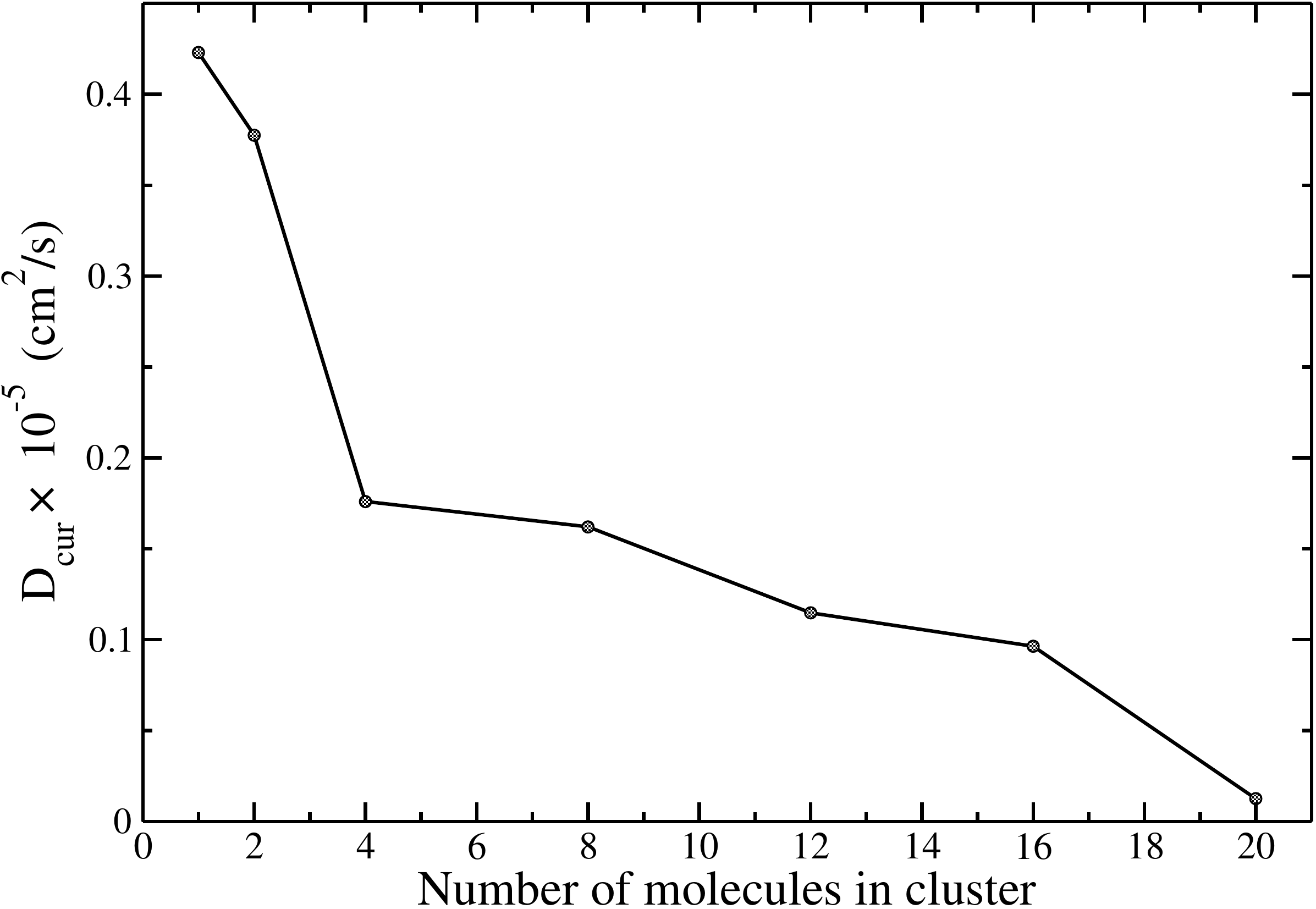}
\caption{\label{fig_msd_cur} (Colour online)
Mean-square displacement of the center of mass of a curcumin molecule 
belonging to a cluster with a size $N_\text{cur}$ (left-hand panel).
Self-diffusion coefficient of a curcumin molecule
belonging to a cluster as a function of the number of molecules in the 
cluster $N_\text{cur}$ (right-hand panel).}
\end{center}
\end{figure}

In figure~\ref{fig_msd_cur}a, we show
the MSD functions of the COM for curcumin cluster of different 
sizes $N_\text{cur}$. It is worth noting that after $20$~ns
these functions behave linearly with time. Therefore, the time interval of $20-40$~ns 
was used to estimate the self-diffusion coefficients
of clusters, $D_\text{cur}$, from the presented MSD functions by using the Einstein relation,
\begin{equation}
D_i =\frac{1}{6} \lim_{t \rightarrow \infty} \frac{\rd}{\rd r} \vert {\bf r}_i(\tau+t)-{\bf r}_i(\tau)\vert ^2,
\end{equation}
where $i$ refers to the COM of a curcumin molecule belonging to a cluster,
and $\tau$ is the time origin. 

One can see that the self-diffusion for $N_\text{cur}=2$
is somewhat smaller ($0.378\cdot10^{-5}$~cm$^{2}$/s) than for the case of single 
curcumin molecule in water ($0.423\cdot10^{-5}$~cm$^{2}$/s) obtained
by us earlier in \cite{Pat-2017}.
However, it drastically decreases to $D_\text{cur}=0.176\cdot10^{-5}$~cm$^{2}$/s for $N_\text{cur}=4$, 
afterwards  it falls  down to $D_\text{cur}=0.096349\cdot10^{-5}$~cm$^{2}$/s for $N_\text{cur}=16$. 
Finally, the self-diffusion coefficient
reaches a very small value $D_\text{cur}=0.012496\cdot10^{-5}$~cm$^{2}$/s for $N_\text{cur}=20$,
figure~\ref{fig_msd_cur}b.
Apparently, a single curcumin molecule and the ``dimer'' aggregate find enough space 
for translational motion in water. 
By contrast, the translation motion of a molecule belonging to a larger cluster becomes
essentially hindered. Finally, in a big cluster, the curcumin molecules become 
almost ``frozen''.

Alternatively, it would be of interest to perform calculations of dynamical properties
by using the velocity auto-correlation functions (VACFs). This method would permit to obtain
the self-diffusion coefficient as well, and to compare the results with the predictions from the MSD.
Most importantly, the VACF calculations provide an ampler set of dynamic properties,
such as relaxation times and vibrational spectra. These issues will be discussed in  
a separate work.

\subsection{On the hydration of clusters}

One of important aspects of the self-assembly of curcumin particles in water is the hydration of
clusters. Cluster formation intuitively should be accompanied by the process in which the
water molecules are expelled from the cluster body. On the other hand, the cluster surface 
formed due to self-assembly should be geometrically and energetically heterogeneous. Thus,
the structure of such an interface can be quite complex. 

\begin{figure}[!t]
  \begin{center}
  \includegraphics[width=0.45\textwidth,angle=0,clip=true]{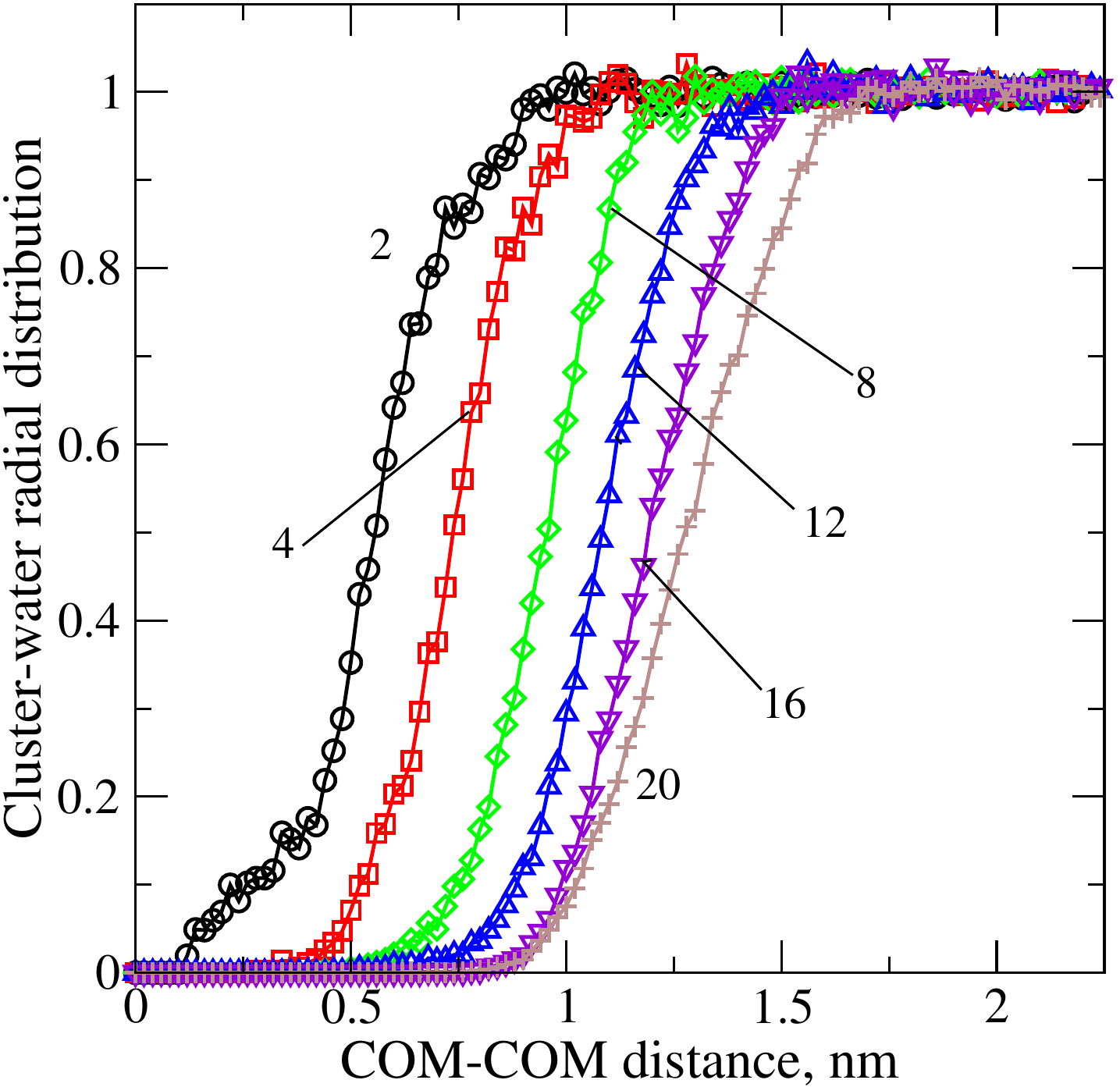}
  \caption{\label{fig_cur_water}(Colour online) Radial distribution function of water molecules around a curcumin cluster 
with respect to its COM.
}
\end{center}
\end{figure}

In order to get insight into this kind of interfaces, we first construct  a radial distribution of 
water oxygens around the COM of a cluster of curcumin species. As expected, the water molecules
are expelled from the cluster body, figure~\ref{fig_cur_water}. The functions constructed for systems
with a different number of curcumins saturate to unity 
at a distance dependent on $N_\text{cur}$. The width of the interface is
diffuse indicating a certain degree of permeation of water molecules into a cluster. 
Moreover, it is difficult to unequivocally establish the trends of changes of the width of this interface.
Apparently, in all the cases, rate of changes of this function is similar, though in the case of a larger
cluster. Namely, if $N_\text{cur}$ changes from 16 to 20, the interface width becomes wider. 

\begin{figure}[!t]
  \begin{center}
  \includegraphics[width=0.4\textwidth,angle=0,clip=true]{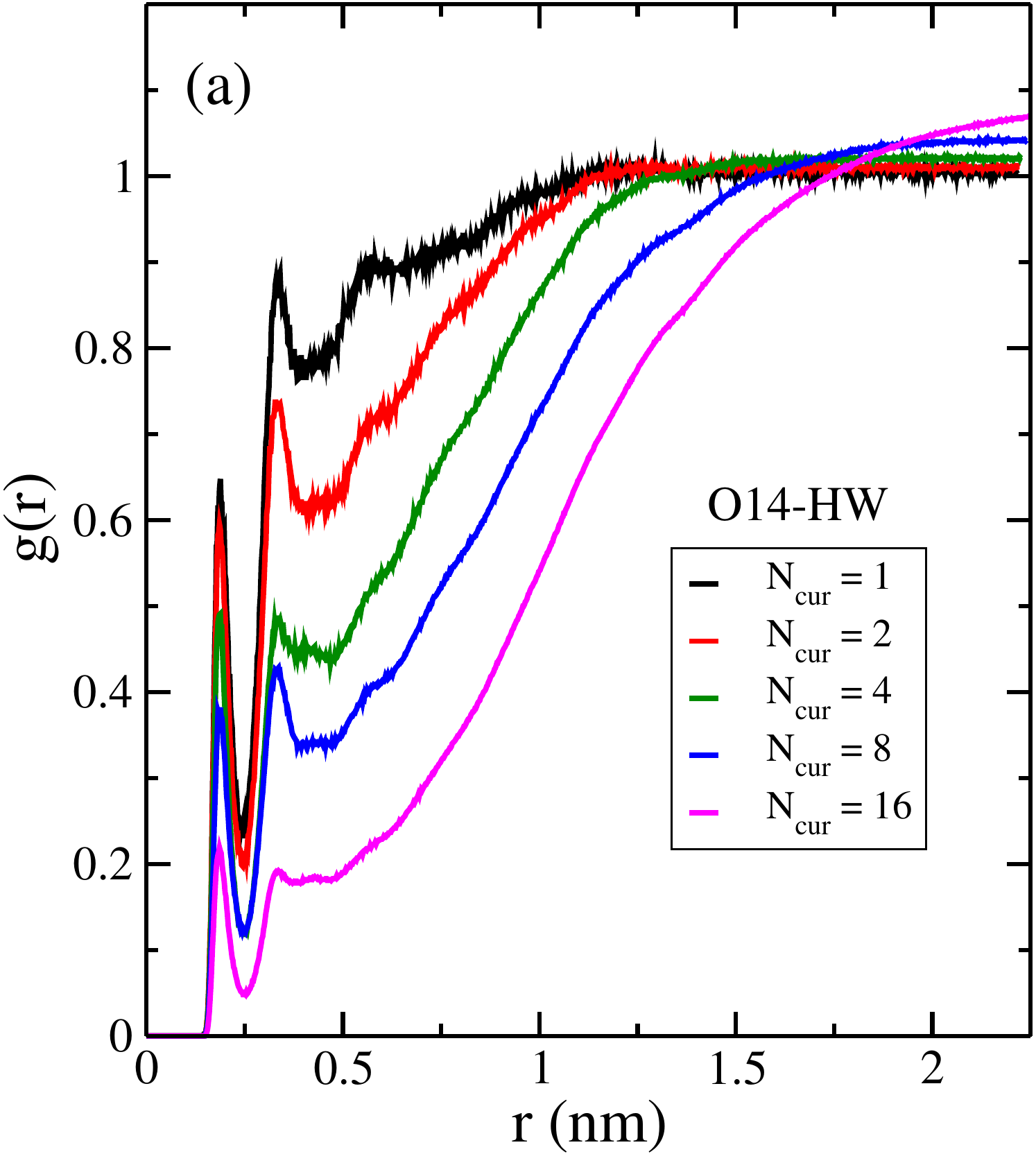}
  \includegraphics[width=0.4\textwidth,angle=0,clip=true]{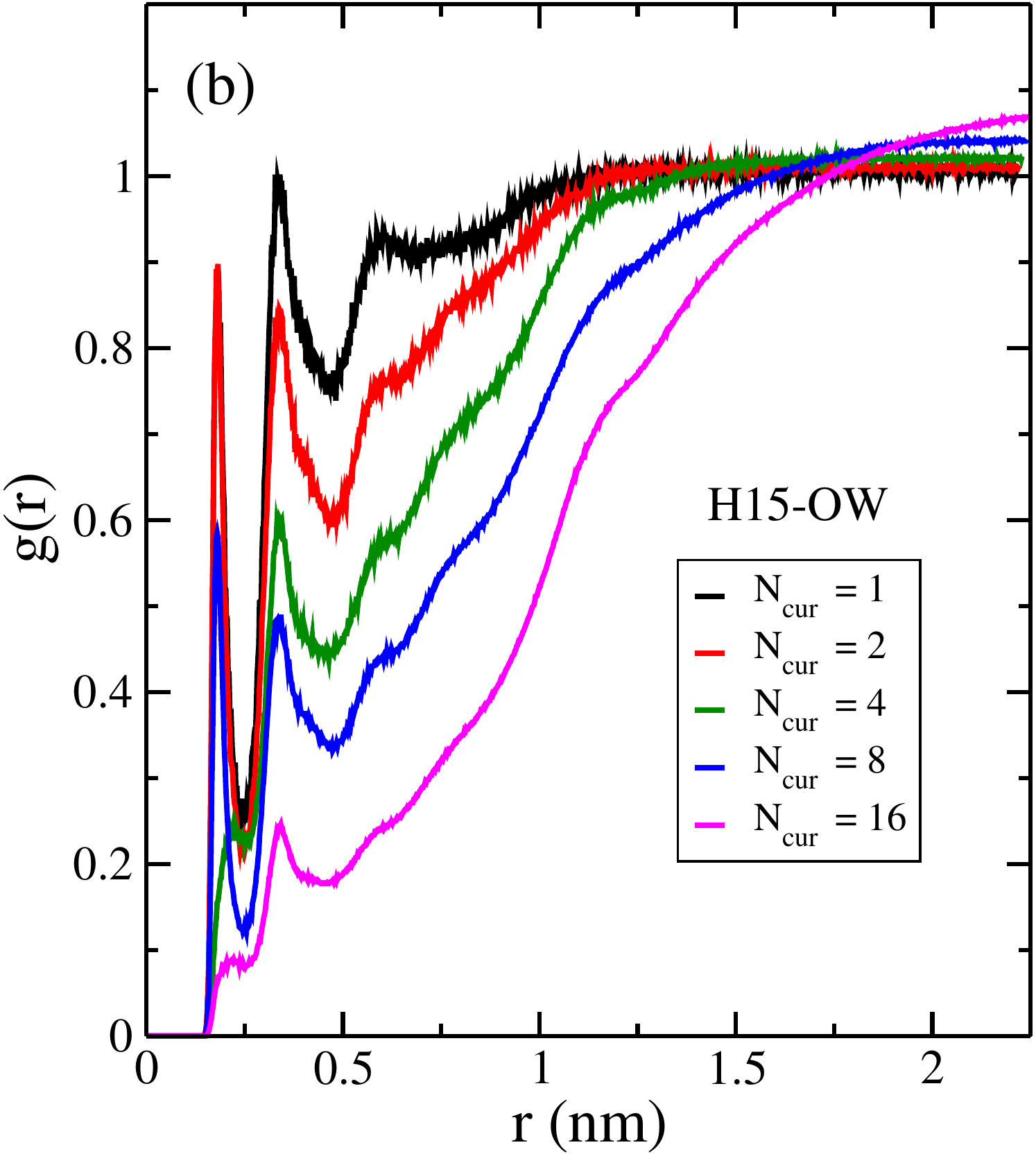}
  \caption{\label{fig_pdf_O14}(Colour online) Pair distribution function O14-HW (left-hand panel) and H15-OW (right-hand panel).
}
\end{center}
\end{figure}

\begin{figure}[!t]
   \begin{center}
   \includegraphics[width=0.4\textwidth,angle=0,clip=true]{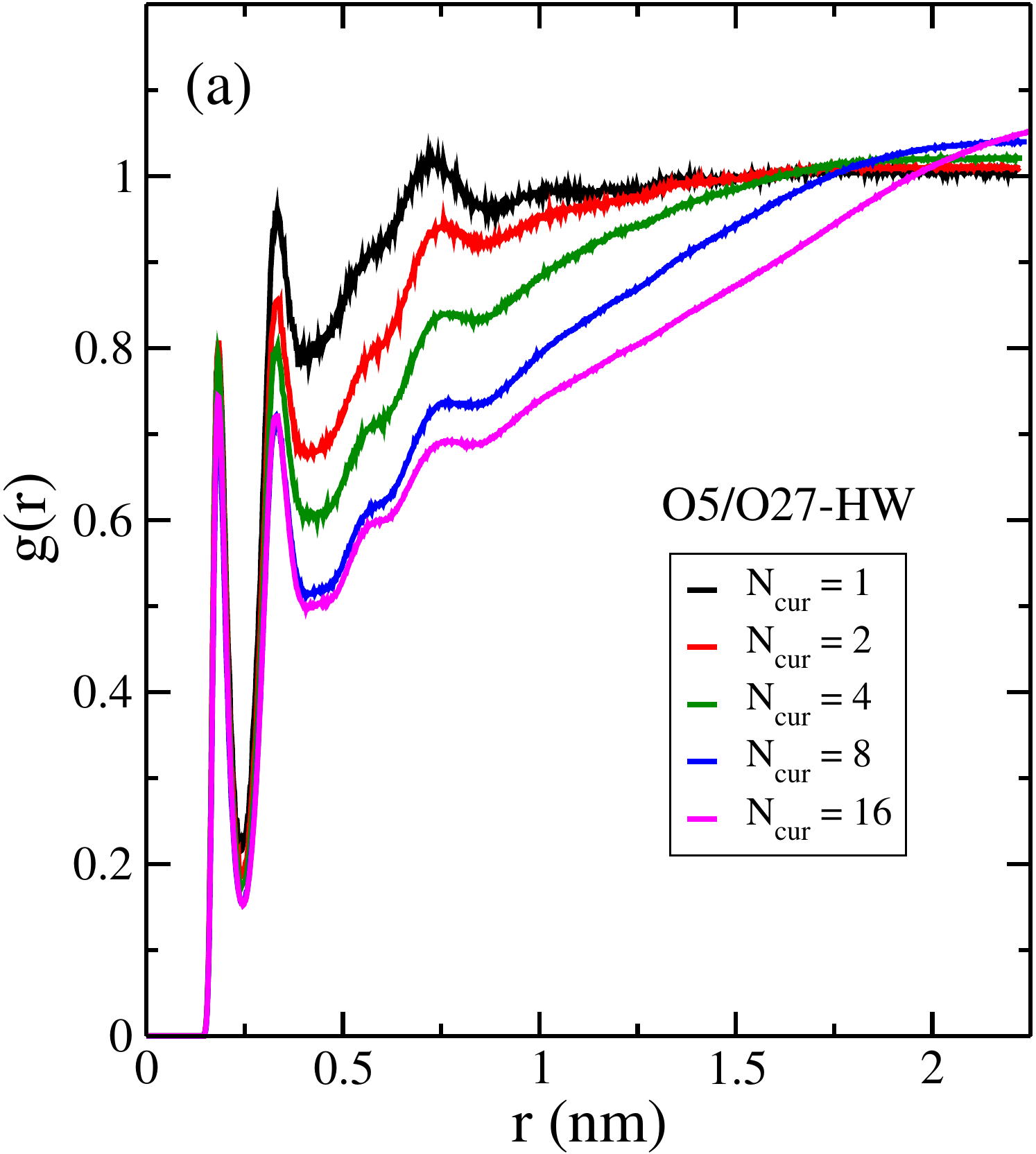}
   \includegraphics[width=0.4\textwidth,angle=0,clip=true]{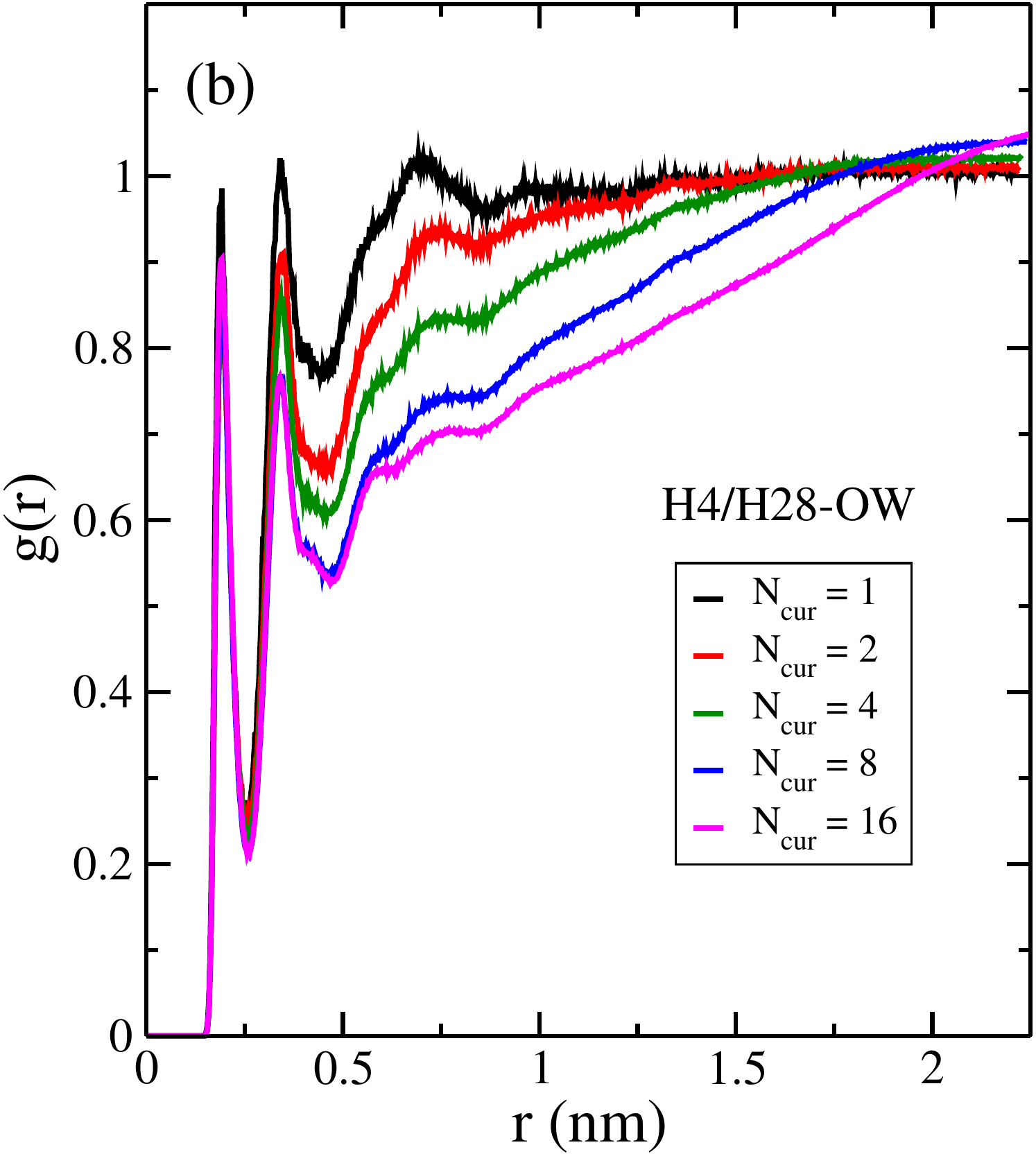}
\caption{\label{fig_pdf_O5}(Colour online) Pair distribution function O5/O27-HW (panel a) and H4/H28-OW (panel b).
}
 \end{center}
\end{figure}

In order to get insights into the structure of the interface formed upon the assembly of curcumin species, we 
 constructed a set of radial distribution functions for different atoms of the curcumin molecule
and water. They are shown in figure~\ref{fig_pdf_O14} and figure~\ref{fig_pdf_O5}.  First, we picked up 
the oxygen and hydrogen atoms, O14 and H15 respectively, of a curcumin molecule as a reference and built up
the radial distribution of atoms of water molecules.  
The corresponding functions, for e.g., $N_\text{cur}=16$, reach saturation
for $r$ larger than $\approx$ 1.6 nm, figure~\ref{fig_pdf_O14}. This estimate agrees with the observations coming
from $g(r)$ in the previous figure~\ref{fig_cur_water}. 
This behavior  could be anticipated, because O14 and H15 atoms are
close to the COM of a molecule and presumably are situated in the cluster interior. It is worth mentioning 
that a single curcumin molecule is quite hydrophobic, i.e., water molecules do not like to approach
O14 and H15, as it follows from the black curves in figure~\ref{fig_pdf_O14}. However, 
as a manifestation of cooperativity in the case of even a small cluster, 
this fragment of a curcumin molecule becomes much stronger hydrophobic.

Next, we picked up oxygen and hydrogen atoms, O5 and H4 respectively, of a curcumin molecule
and plot radial distribution of atoms of water molecules, figure~\ref{fig_pdf_O5}. These atoms are far from
the center of mass of a single curcumin molecule.  Moreover, according to the snapshots and other
structural indicators discussed above, these atoms should be situated on the external surface of a cluster.
Consequenctly, we observe that the cooperative effect of self-assembly of curcumins into a cluster
is not very strong at small inter-particle separations. A certain amount of water molecules prefers to 
locate quite close to O5 and H4 even if the cluster is formed, 
quite similarly to a single 
curcumin molecule in water. 
Moreover, the shape of the functions with multiple peaks in figure~\ref{fig_pdf_O5} 
indicate the existence of water structure close to O5 and H4. Thus, on average the degree of hydrophobicity
of these fragments of curcumin molecules is much less pronounced, in comparison to the fragments involving
O14 and H15.

\begin{figure}[!t]
   \begin{center}
   \includegraphics[width=0.4\textwidth,angle=0,clip=true]{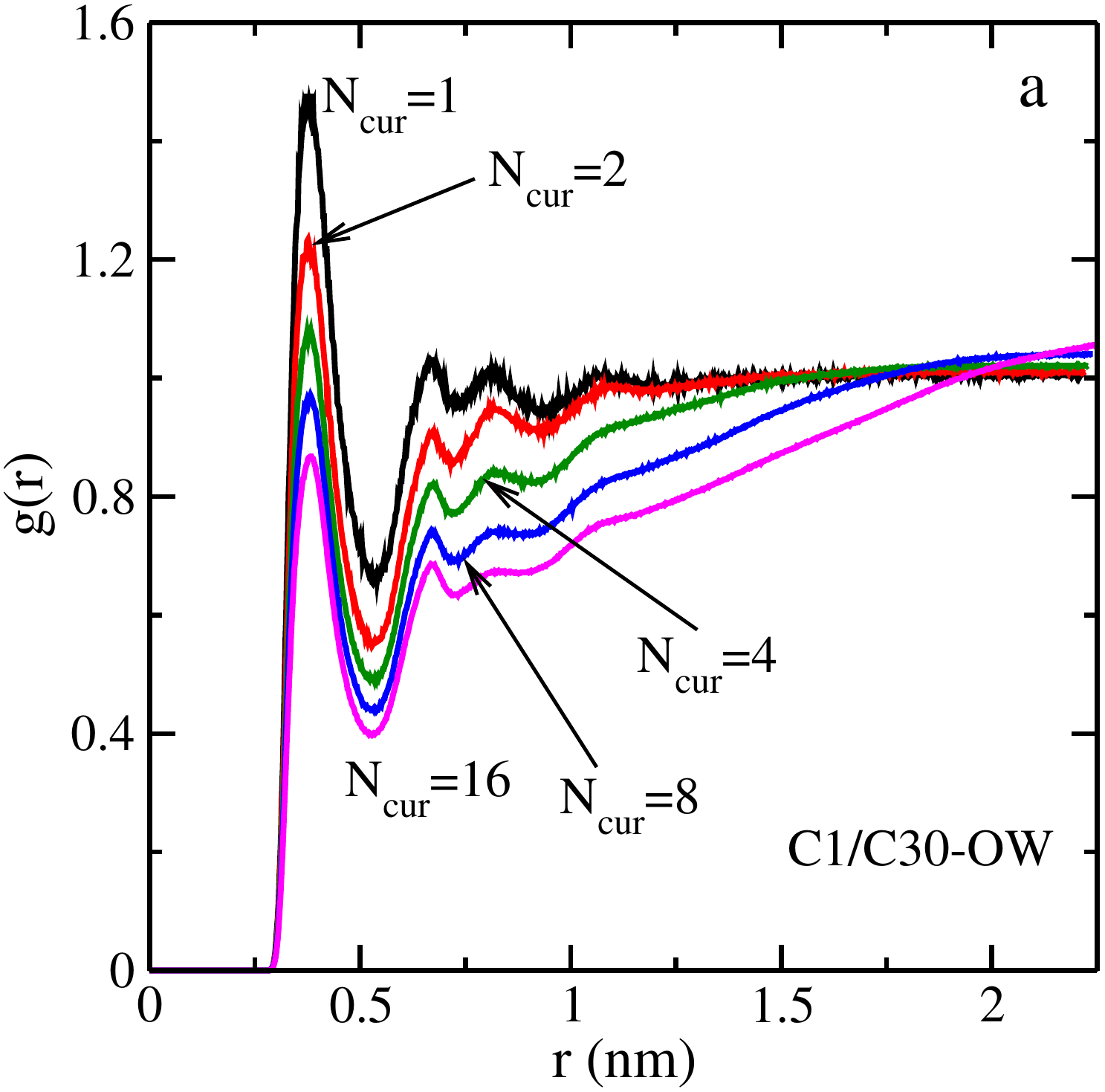}
   \includegraphics[width=0.4\textwidth,angle=0,clip=true]{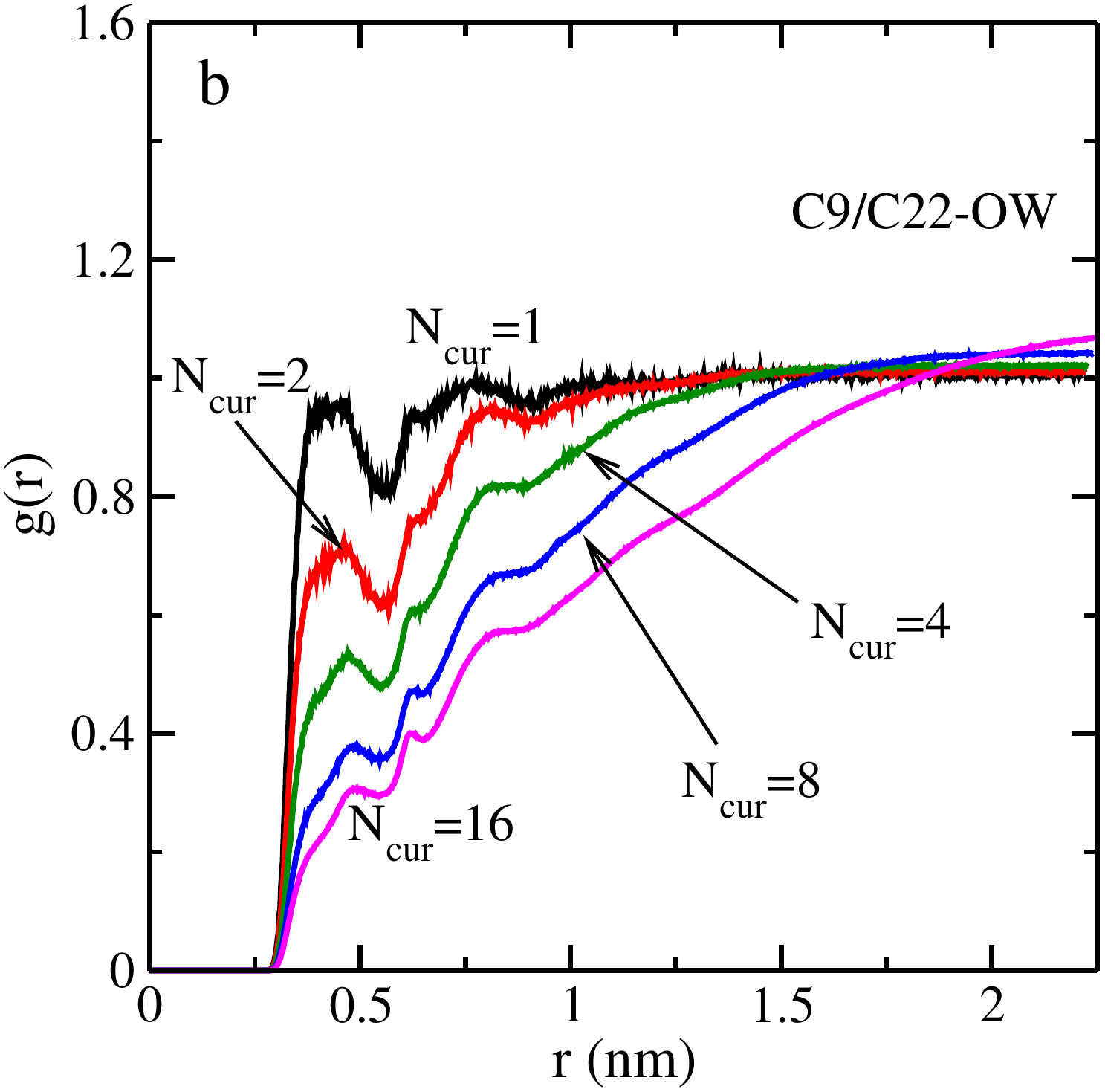}
	\caption{\label{fig_pdf_O6}(Colour online) Pair distribution function  C1/C30-OW (panel a) and C9/C22-OW (panel b).
}
 \end{center}
\end{figure}

In addition, we explored the distribution of water close to the carbon groups of the phenyl rings.
The results are shown in figure~\ref{fig_pdf_O6}.  
Apparently, this distribution of water molecules is strongly affected by the
formation of a cluster. The entire curve yielding this kind of $g(r)$ moves down 
strongly, if the cluster size increases, indicating cooperative hydrophobicity. 
The shape of radial distribution in the present case (for a large cluster) is quite similar
to what we observed in figure~\ref{fig_pdf_O5}. 
However, the water structure close to C1 (or equivalently close to C30)
is less pronounced or smoother, in comparison to the structure close to~O5.

In summary, less hydrophobicity of the surface of a cluster for the systems of this study
can be attributed to O5, H4,  or equivalently to O27, H28 fragments of the curcumin molecules.

\section{Summary and conclusions}

To conclude, in this work we have presented a very detailed description of the properties
of solutions consisting of a different number of curcumin molecule
(from 2 to 16) in 3000 molecules of water. The home-made, non-polarizable OPLS-UA model 
for curcumin molecule was used. 
On the other hand, water is considered by using the SPC-E model. Geometric
combination rules were applied. The simulations were performed by using
isobaric-isothermal conditions. Our principal issue was in obtaining and analysing
 the structural aspects of self-assembly of curcumin molecules in an aqueous medium.

General trends of our findings are in a qualitative agreement with the recent
 similar kind of systems performed in the laboratory of Bagchi~\cite{Hazra-2014}
by using a different model of curcumin molecules. However, we offer a wider set
of properties in comparison to that work. Namely, we analyzed 
various orientational descriptors for the distribution of curcumins in a cluster,  
the radius of gyration of a cluster, order parameter and diffusion coefficient.
The cluster-water interface is characterized by several radial distribution 
functions that indicate its geometric and energetic heterogeneity and an overall
hydrophobicity. 

Still, similarly to \cite{Hazra-2014}, we were unable to find an appropriate experimental
setup to verify our predictions from computer simulations versus experiment.
Apparently, this can be done in future, by adding a certain amount of a co-solvent that
should lead to disaggregation of clusters. Specifically, the systems consisting of curcumin species
in water-ethanol solvent of variable composition
were studied very recently using molecular dynamics computer simulations 
and experimental methods~\cite{pereira}. The study made use of determining the so-called
critical water aggregation percentage (CWAP) to delimit the monomeric form of curcumin species 
from aggregates.
Experimentally, the CWAP, i.e., the percentage of water below which curcumin is in 
its  monomeric form predominantly,  
was evaluated from the analyses of the effect of electronic absorption
spectra and of fluorescence emission spectra on the solvent composition, see e.g., the 
discussion of figures~2 and 3 of \cite{pereira}. 
This quantity may provide good test of the quality of the curcumin model of the present study.

Several interesting and important issues, even within the present stage of modelling, require 
additional investigation. 
Namely, a  more detailed description of the solute - solvent interface would be desirable
in perspective. The properties of a solvent and of a co-solvent, if present, 
around  a biomolecule determine its conformations and affect trends to
form clusters. Evidently, the ``direct'' interaction between such complex solutes is of importance
in the latter aspect as well. Moreover, at the level of a more 
sophisticated modelling, one may attempt to elucidate the role of the 
polarizability of solute molecules.
At present, for the system of our interest, it is difficult to profoundly discern and analyze 
each type of the effects. 
Interesting discussion of some of these issues for a simpler, dimethylsulfoxide-water solutions
was given in \cite{bachmann}.

Concerning the formation of clusters of curcumin molecules, one may attempt to take advantage
of the system with two curcumin molecules dispersed in a solvent. Then, along the line of previous studies
of simpler molecular fluids, one may focus on the evaluation of the potential of the mean
force (PMF). The software of the present study permits to obtain the PMF for the center of mass
of curcumin species. Then, the second virial coefficient at a given thermodynamic
conditions can be calculated. The molecular clustering phenomena can be analyzed in terms
of the virial coefficients, as it was proposed originally in \cite{woolley} 
and implemented for water in the laboratory of Kofke \cite{kofke1,kofke2}. This kind of
methodology is of interest in a wider context and for the systems of the present work 
as well, see e.g., \cite{harding}.
Some of these issues are under study in our research groups.

	\section*{Acknowledgement}
O.P. is grateful to M. Aguilar for technical support of this
work at the Institute of Chemistry of UNAM. Fruitful discussions with Dr. Manuel Soriano 
at the early stage of this project  are gratefully acknowledged. T.P. acknowledges allocation of
computer time at the cluster of ICMP of the National Academy of Science of Ukraine and
Ukrainian National Grid.

\newpage
\ukrainianpart

\title{Структурні аспекти кластерування молекул куркуміну в воді. Комп'ютерне моделювання методом молекулярної динаміки}
\author{Т. Пацаган\refaddr{label1}, О. Пізіо\refaddr{label2}}
\addresses{
	\addr{label1} Інститут фізики конденсованих систем Національної академії наук України, \\вул. Свєнціцького, 1, 79011 Львів, Україна
	\addr{label2} Інститут Хімії, Нацiональний автономний унiверситет Мексики, м. Мехiко, Мексика
}
%
%
%

\makeukrtitle

\begin{abstract}
	Ми досліджуємо кластерування молекул куркуміну у воді, використовуючи модель OPLS-UA для енольної форми куркуміну 
	(J. Mol. Liq., \textbf{223}, 707, 2016) та модель SPC-E для води. З цією метою проведено комп'ютерне моделювання 
	розчинів 2, 4, 8, 12, 16 та 20 молекул куркуміну в 3000 молекулах води із використан\-ням методу молекулярної динаміки. 
	Розраховано радіальні розподіли центрів мас молекул куркуміну та проаналізовано значення біжучих координаційних чисел. 
	Прослідковано формування кластерів з часом. Отримано опис внутрішньої структури молекул у кластері за допомогою радіальних розподілів окремих елементів молекули куркуміну, орієнтаційних дескрипторів, параметра порядку та радіусу гірації. Розраховано коефіцієнт самодифузії кластерованих молекул куркуміну. Детально описано розподіл молекул води навколо кластерів. Виконано порівняння наших результатів з результатами комп’ютерного моделювання інших авторів. 
	Обговорюється можливість зв'язку передбачень, отриманих для нашої моделі, та експериментально спостережуваних даних. %
	
	\keywords куркумін, модель об’єднаних атомів, молекулярна динаміка, вода, кластери
	
\end{abstract} 
\end{document}